\documentclass[a4paper,11pt]{article}
\pdfoutput=1 

\usepackage{jcappub} 

\usepackage[T1]{fontenc} 

\usepackage[utf8]{inputenc}
 
\usepackage{enumerate} 
\usepackage{afterpage} 
\usepackage{amsmath} 
\usepackage{ulem}
\RequirePackage{lineno}

\usepackage[usenames,dvipsnames]{xcolor}

\definecolor{forestgreen}{rgb}{0.13, 0.55, 0.13}
\definecolor{cssgreen}{rgb}{0.07, 0.53, 0.03}

\newcommand{\leff}{\ensuremath{\mathcal{L}_{\rm eff}}}		

\DeclareTextFontCommand{\emph}{\it}	

\title{DARWIN: towards the \\ ultimate dark matter detector}

\collaboration{DARWIN collaboration}

\author[a]{J.~Aalbers,}
\author[b,n]{F.~Agostini,}
\author[c]{M.~Alfonsi,}
\author[r]{F.D.~Amaro,}
\author[d]{C.~Amsler,}
\author[e]{E.~Aprile,}
\author[f]{L.~Arazi,}
\author[g]{F.~Arneodo,}
\author[h]{P.~Barrow,}
\author[h]{L.~Baudis,}
% ...

\affiliation[a]{Nikhef and the University of Amsterdam, Netherlands}
\affiliation[b]{Department of Physics and Astrophysics, University of Bologna and INFN-Bologna, Italy}
\affiliation[c]{Institut f\"ur Physik \& Exzellenzcluster PRISMA, Johannes Gutenberg-Universit\"at Mainz, Germany}
\affiliation[d]{Albert Einstein Center for Fundamental Physics, Universit\"at Bern, Switzerland}
\affiliation[e]{Physics Department, Columbia University, New York, NY, USA}
\affiliation[f]{Department of Particle Physics and Astrophysics, Weizmann Institute of Science, Rehovot, Israel}
\affiliation[g]{New York University Abu Dhabi, United Arab Emirates}
\affiliation[h]{Physik-Institut, Universit\"at Z\"urich, Switzerland}
% ...
\affiliation[n]{INFN-Laboratori Nazionali del Gran Sasso and Gran Sasso Science Institute, L'Aquila, Italy}
% ...
\affiliation[r]{Department of Physics, University of Coimbra, Portugal}
% ...
    
\emailAdd{lior.arazi@weizmann.ac.il}
\emailAdd{laura.baudis@uzh.ch}
\emailAdd{amos.breskin@weizmann.ac.il}
\emailAdd{decowski@nikhef.nl}
\emailAdd{marc.schumann@lhep.unibe.ch}

\abstract{ 
DARk matter WImp search with liquid xenoN (DARWIN\footnote{\tt www.darwin-observatory.org}) will be an experiment   for the direct detection of dark matter using a multi-ton liquid xenon time projection chamber at its core. Its primary goal will be to explore the  experimentally accessible parameter space for Weakly Interacting Massive Particles (WIMPs) in a wide mass-range, until neutrino interactions with the target become an irreducible background. The prompt scintillation light and the charge signals induced by particle interactions in the xenon will be observed by VUV sensitive, ultra-low background photosensors. Besides its excellent sensitivity to WIMPs  above a mass of 5\,GeV/$c^2$,  such a detector with its large mass, low-energy threshold and ultra-low background level will also be sensitive to other  rare interactions. It will search for solar axions, galactic axion-like particles and the neutrinoless double-beta decay of $^{136}$Xe, as well as measure the low-energy solar neutrino flux with $<$1\% precision, observe coherent neutrino-nucleus interactions, and detect galactic supernovae.  We present the concept of the DARWIN detector and discuss its physics reach, the main sources of backgrounds and the ongoing detector design and R\&D efforts.}


\begin{document}
\maketitle
\flushbottom

\section{Introduction}

Astronomical and cosmological observations reveal that the vast majority of the matter and energy content of our universe is invisible -- or dark -- and interacts neither strongly nor electromagnetically with ordinary matter. Results from the Planck satellite~\cite{Ade:2013zuv} show that about 68\% of the overall budget is dark energy, leading to the observed accelerated expansion of the cosmos. Another 27\% is composed of dark matter, a yet-undetected form of matter whose presence is needed to explain the observed large-scale structures and galaxies. While dark matter interacts gravitationally with baryonic matter, any additional interactions, if existing, must be very weak with extremely small cross sections~\cite{Goodman:1984dc}. Because the standard model of particle physics does not accommodate dark matter, the observationally-driven need for its existence is one of the strongest indications for physics beyond the standard model. The direct detection and subsequent characterisation of dark matter particles is, therefore, one of the major experimental challenges of modern particle and astroparticle physics~\cite{Bertone:2004pz,Baudis:2016qwx}.

Many theories beyond the standard model  predict viable candidates; one particular class, receiving the attention of most current and planned experiments, is that of Weakly Interacting Massive Particles (WIMPs)~\cite{Jungman:1995df,Bertone:2004pz}. Worldwide, more than a dozen experiments are prepared to observe low-energy nuclear recoils induced by galactic WIMPs in ultra-sensitive, low-background detectors~\cite{Baudis:2012ig, Schumann:2015wfa, Baudis:2015mpa,Undagoitia:2015gya}. Since the predicted WIMP masses and scattering cross sections are model-dependent and essentially unknown, these searches must cover a vast parameter space~\cite{Arrenberg:2013rzp,Strege:2014ija}. Most promising are detectors based on liquefied noble gas targets such as liquid xenon (LXe) or liquid argon (LAr). This technology is by now well-established and  can be scaled up to ton-scale, homogeneous target masses~\cite{Chepel:2012sj,Schumann:2014uva,Baudis:2014naa}, taking data over several years. 

Two detector concepts are in use. The first uses a single-phase noble-liquid WIMP target, surrounded by photosensors to record the emitted scintillation light. Examples are the XMASS detector, operating a 850\,kg total LXe target~\cite{Abe:2013tc}, as well as DEAP-3600~\cite{Boulay:2012hq} and miniCLEAN~\cite{Rielage:2014pfm}, large LAr detectors currently under commissioning. The LAr instruments employ the powerful rejection of electronic recoil background based on pulse shape discrimination (PSD)~\cite{ref::psd}. With a 3600\,kg LAr target, the larger detector DEAP-3600 aims at a sensitivity of $\sim$$1\times10^{-46}$\,cm$^2$ for spin-independent WIMP-nucleon interactions at a WIMP mass of 100\,GeV/c$^2$. 

The second concept is based on dual-phase noble gas time projection chambers (TPCs), where the prompt scintillation light~(S1) and the delayed proportional scintillation light signal from the charge~(S2) are measured.  Both signals are employed for a precise reconstruction of the event vertex and, thus, to suppress backgrounds by rejection of  multiple-scatter interactions, as WIMPs are expected to interact only once. The charge-to-light ratio, S2/S1, is exploited to separate the expected signal, namely nuclear recoils (NR), from the dominant electronic recoil (ER) background. 

TPCs filled with LXe were pioneered by the ZEPLIN~\cite{Akimov:2011tj,Akimov:2010zz} and XENON10~\cite{Angle:2007uj,Aprile:2010bt} collaborations. The XENON100 experiment~\cite{Aprile:2011hi,Aprile:2011dd}, a TPC with a 62\,kg active target, has reached its sensitivity goal and excluded spin-independent WIMP-nucleon cross sections above $2 \times10^{-45}$\,cm$^2$ at a WIMP mass of 55\,GeV/$c^2$~\cite{Aprile:2012nq}. These constraints were superseded by the results from the LUX collaboration~\cite{Akerib:2012ys,Akerib:2015rjg}, which operates a 250\,kg TPC and excludes spin-independent WIMP-nucleon scattering cross sections above $4 \times 10^{-46}$\,cm$^2$ at 33\,GeV/c$^2$. The second phase of PandaX has published first result from its run with a 500\,kg active LXe target~\cite{Tan:2016diz}. 
Liquid argon dual-phase TPCs were pioneered by WArP~\cite{Benetti:2007cd} and ArDM~\cite{Amsler:2010yp}.
In addition to the S2/S1~discrimination, they also exploit the considerably more powerful PSD rejection of ER background~\cite{ref::psd}. DarkSide-50, using 50\,kg of active LAr mass, has presented first results from a low radioactivity run~\cite{Agnes:2015ftt}. The experiment reduces its target radioactivity by using underground argon in which the radioactive $^{39}$Ar is depleted by a factor of 1.4$\times$10$^3$ with respect to atmospheric argon.

\begin{figure}[h!]
\includegraphics*[width=0.50\textwidth]{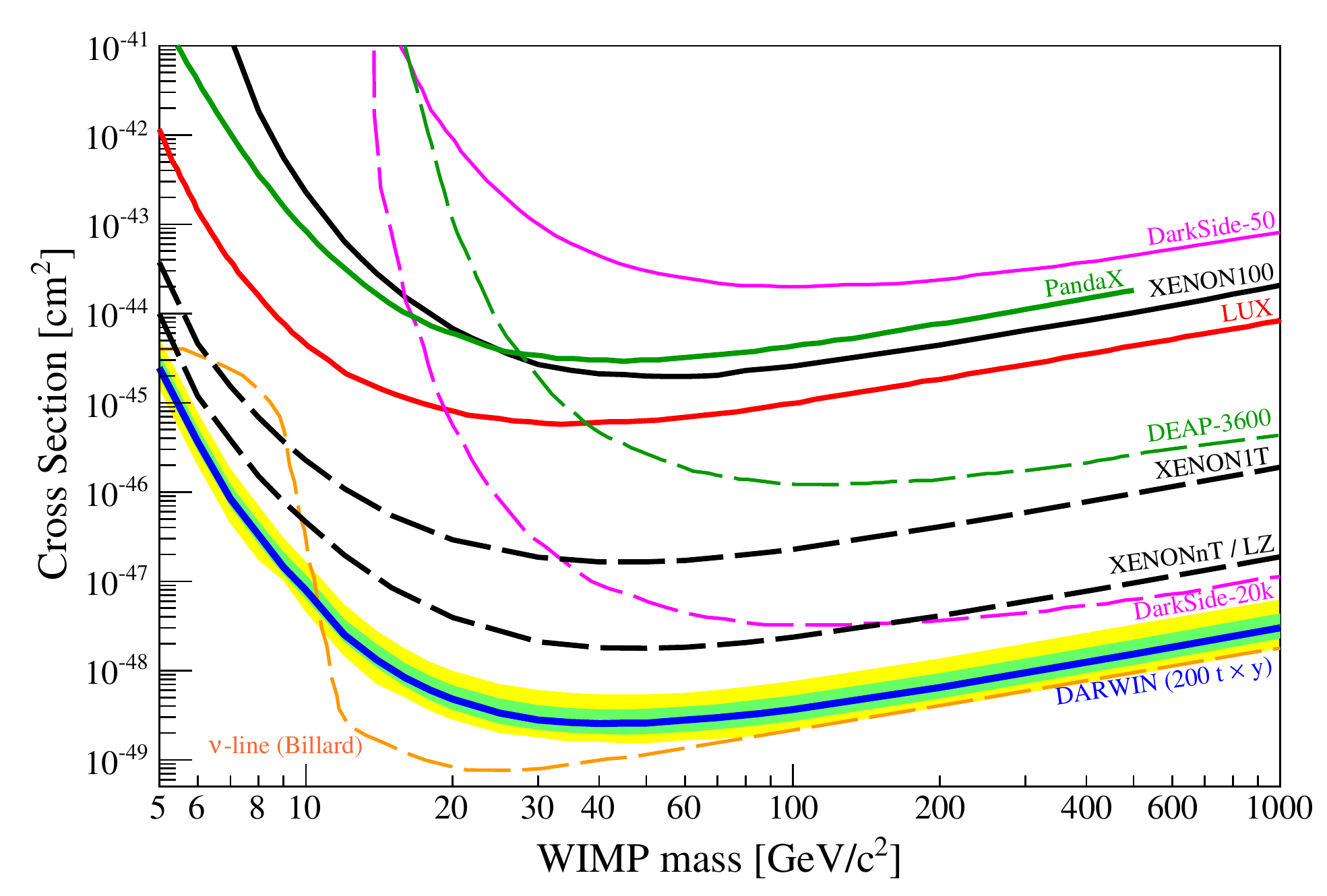}    
\includegraphics*[width=0.50\textwidth]{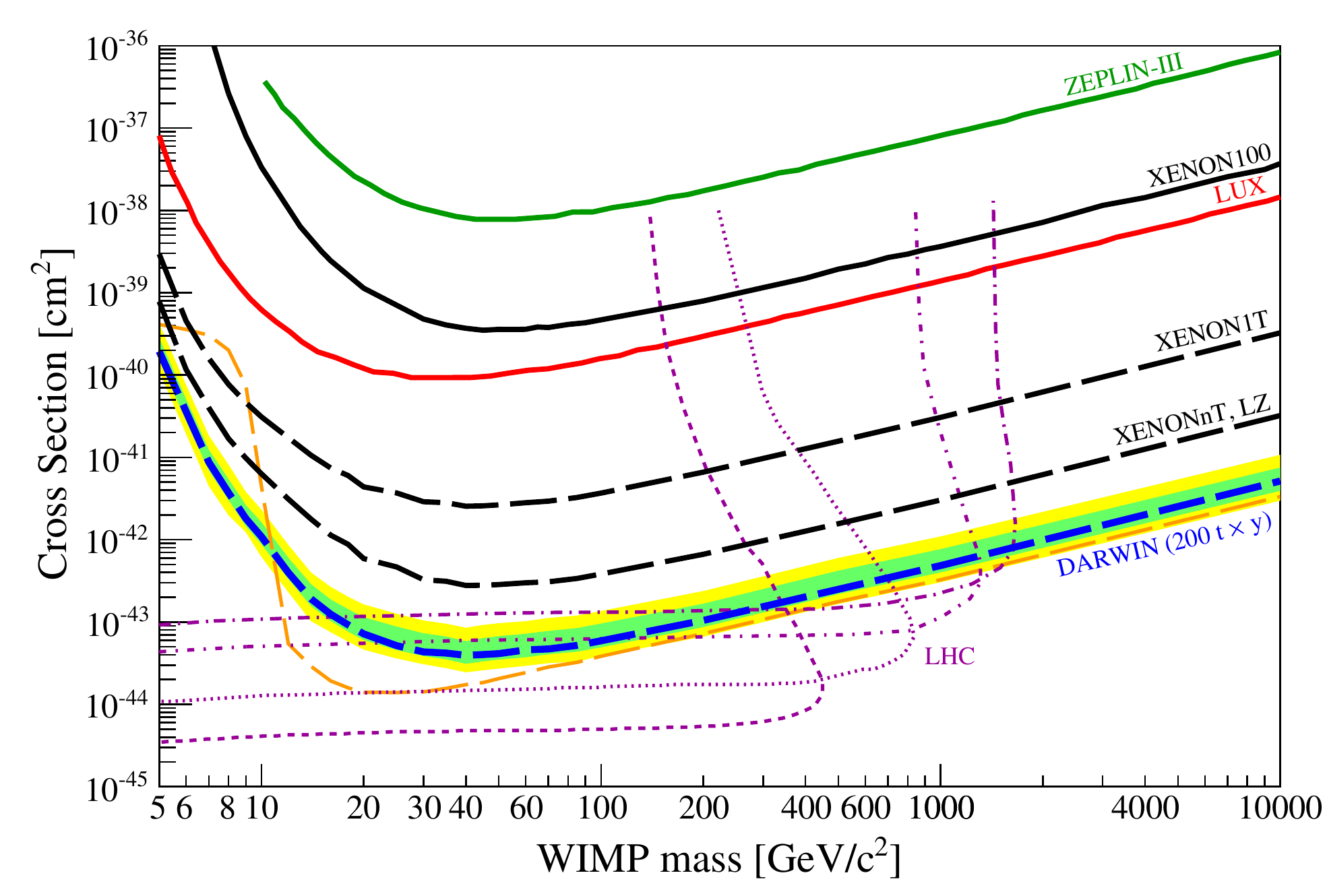} 
\caption{\small {\textbf{(left)} The attainable bound on spin-independent WIMP nucleon cross-section as a function of the WIMP mass of present and future nobel liquid detectors. Shown are upper limits from PandaX-II~\cite{Tan:2016diz},  DarkSide-50 \cite{Agnes:2015ftt}, XENON100~\cite{Aprile:2012nq}, and LUX~\cite{Akerib:2015rjg} as well as the sensitivity projections for DEAP3600~\cite{Boulay:2012hq}, XENON1T~\cite{Aprile:2015uzo}, XENONnT~\cite{Aprile:2015uzo}, LZ~\cite{Akerib:2015cja}, DarkSide-20k~\cite{ref::ds_20k} and DARWIN~\cite{Schumann:2015cpa}, for which we also show the 1-$\sigma$ (yellow band) and 2-$\sigma$ (green band) regions.  DARWIN is designed to probe the entire parameter region for WIMP masses above $\sim$5\,GeV/c$^2$, until the neutrino background ($\nu$-line, dashed orange~\cite{Billard:2013qya}) will start to dominate the recoil spectrum. \textbf{(right)} Upper limits on the spin-dependent  WIMP-{\it neutron} cross section of ZEPLIN-III~\cite{Akimov:2011tj}, XENON100~\cite{Aprile:2013doa} and LUX~\cite{Akerib:2016lao} as well as projections for XENON1T, XENONnT, LZ and DARWIN. DARWIN and the high-luminosity LHC will cover a common region of the parameter space. The 14\,TeV LHC limits for the coupling constants $g_\chi$\,$=$\,$g_q$\,$=$\,0.25, 0.5, 1.0, 1.45 (bottom to top) are taken from \cite{Malik:2014ggr}. The LHC reach for spin-independent couplings is above 10$^{-41}$\,cm$^2$. Argon has no stable isotopes with non-zero nuclear spins and is thus not sensitive to spin-dependent couplings. Figures updated from~\cite{Schumann:2015cpa}, using the xenon response model to nuclear recoils from~\cite{Akerib:2015rjg} for the DARWIN sensitivity.}}
\label{fig::limits}
\end{figure}

Probing lower cross sections at WIMP masses above a few GeV/$c^2$ requires larger detectors.  XENON1T, the current phase in the XENON collaboration programme, aims to reach spin-independent cross sections of $1.6\times10^{-47}$\,cm$^2$ after 2~years of continuous operation of its 2\,t LXe target~\cite{Aprile:2015uzo}. The next phase, XENONnT, to be designed and constructed during XENON1T operation, will increase the sensitivity by another order of magnitude, assuming 20 t$\times$y exposure~\cite{Aprile:2015uzo}.  A similar sensitivity is sought by LUX-ZEPLIN (LZ), the next phase in the LUX programme, which plans to operate a  7\,t LXe detector  with an additional scintillator veto to suppress the neutron background~\cite{Akerib:2015cja}. The DarkSide collaboration proposes a 20\,t LAr dual-phase detector, with the goal to reach  $9\times10^{-48}$cm$^2$ at 1\,TeV/$c^2$, based on extrapolations of the demonstrated PSD efficiency of the smaller detector~\cite{Alexander:2013hia,ref::ds_20k}.  

The DARk matter WImp search with liquid xenoN (DARWIN) observatory, which is the subject of this article, aims at a $\sim$10-fold increase in sensitivity compared to these projects. Figure~\ref{fig::limits}~(left) summarises the status and expected sensitivities to spin-independent WIMP-nucleon interactions as a function of the WIMP mass for noble liquid detectors including DARWIN. Figure~\ref{fig::limits}~(right) shows the situation for the spin-dependent case, assuming WIMP coupling to neutrons only. DARWIN and the high-luminosity LHC will cover common parameter space~\cite{Malik:2014ggr} in this channel.

This article is structured as follows: After a brief introduction to the DARWIN project in Section~\ref{sec::darwin}, we discuss its reach for several astroparticle and particle physics science channels in Section~\ref{sec:science}. DARWIN's main background sources are introduced in Section~\ref{sec:backgrounds}, followed by a detailed discussion on design considerations and the status of the ongoing R\&D towards the ultimate WIMP dark matter detector in Section~\ref{sec:r_and_d}.

\section{The DARWIN project}
\label{sec::darwin}

DARWIN will be an experiment using a multi-ton liquid xenon TPC, with the primary goal to explore the experimentally accessible parameter space for WIMPs. DARWIN's 50\,t total (40\,t active) LXe target will probe particles with masses above 5\,GeV/$c^2$, and WIMP-nucleon cross sections down to the few $\times10^{-49}$\,cm$^2$ region for masses of $\sim$50\,GeV/$c^2$~\cite{Schumann:2015cpa}.  Should dark matter particles be discovered by existing or near-future experiments, DARWIN will measure WIMP-induced nuclear recoil spectra with high statistics and constrain the mass and the scattering cross section of the dark matter particle~\cite{Pato:2010zk,Newstead:2013pea}. Other physics goals are the real-time detection of solar $pp$-neutrinos with high statistics, detection of coherent neutrino-nucleus interactions, searches for solar axions and galactic axion-like particles (ALPs) and the search for the neutrinoless double beta decay ($0\nu\beta\beta$). The latter would establish whether the neutrino is its own anti-particle, and can be detected via the double beta emitter $^{136}$Xe, which has a natural abundance of 8.9\% in xenon. The facility will also be able to observe neutrinos of all flavours from supernova explosions~\cite{Chakraborty:2013zua}, providing complementary information to large-scale water-Cherenkov or LAr detectors. DARWIN is included in the European Roadmap for Astroparticle Physics (APPEC) and additional national roadmaps.

\begin{figure}[h!]
\begin{minipage}[]{0.47\textwidth}
\begin{center}
\includegraphics*[width=1.\textwidth]{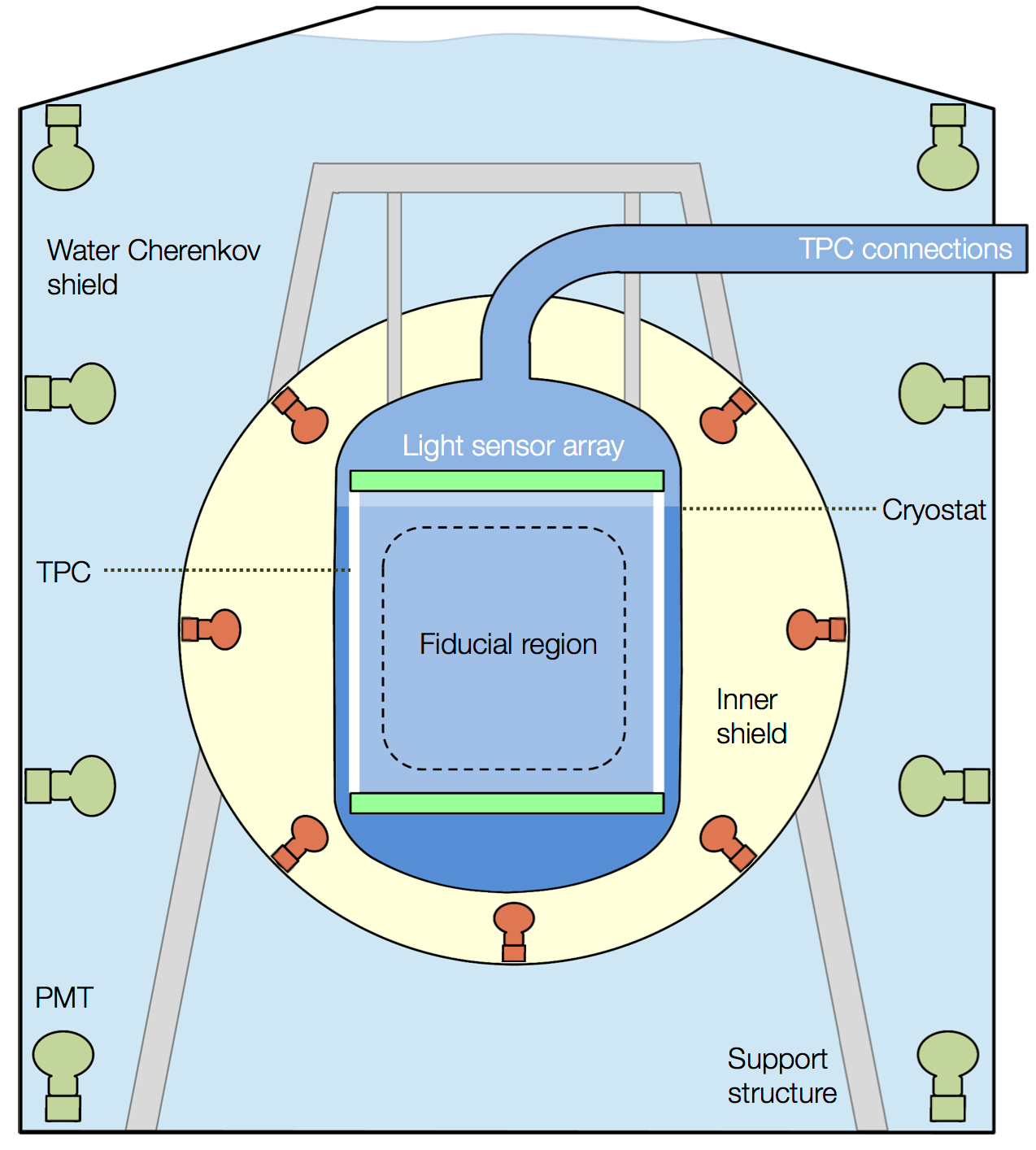}  
\end{center}
\end{minipage}
\hfill 
\begin{minipage}[]{0.47\textwidth}
\vspace{-2.2cm}
{\small \caption{Sketch of the DARWIN detector inside a tank, operated as a water-Cherenkov muon veto. The need for an additional liquid-scintillator neutron veto inside the water shield, as shown in the figure (`inner shield'), is subject to further studies. The dual-phase time projection chamber is enclosed within a double-walled cryostat and contains $40$\,t of liquid xenon (50\,t total in the cryostat). In the baseline scenario, the prompt and delayed VUV scintillation signals, induced by particle interactions in the sensitive  volume, are recorded by two arrays of photosensors installed above and below the liquid xenon target. \label{fig::darwin} }}
\end{minipage}
\end{figure}

The experiment will operate a large volume of liquid xenon in a low-background cryostat, surrounded by concentric shielding structures, as shown schematically in Figure~\ref{fig::darwin}. Future studies will reveal whether a liquid scintillator detector inside the water Cherenkov shield is required for this massive detector. The core of the experiment is a dual-phase TPC  containing the active xenon mass. The high density of liquid xenon,  $\sim$3\,kg/l, results in a short radiation length and allows for a compact detector geometry with efficient self-shielding. The fiducial target mass is not fixed {\sl a priori}: it will be defined during the analysis process, based on the relevant backgrounds and on the studied physics case. A drift field of $\cal O$(0.5)\,kV/cm  across the liquid target is required to drift the electrons from the interaction vertex. This will be achieved by biasing the cathode at the bottom of the TPC  with voltages on the order of $-100$\,kV or above. Large field shaping rings made from oxygen-free high conductivity (OFHC) copper,  optimised for such high voltages, will ensure the field homogeneity.   The main materials to be used for the TPC construction are OFHC copper as a conductor and polytetrafluoroethylene (PTFE) as an insulator, with the latter also acting as an efficient reflector for vacuum ultra-violet (VUV) scintillation light~\cite{ref::ptfe_reflectivity}. The TPC will be housed in a double-walled cryostat made out of stainless steel, titanium or copper, and all the materials are to be selected for ultra-low intrinsic radioactivity. The structure will be suspended from a support frame which will allow for the levelling of the TPC with $\sim$100\,$\mu$m precision once the outer shields and the detector are filled with liquids.

In the baseline scenario, the prompt  and  proportional scintillation signals will be recorded by two arrays of photosensors  installed above and below the target. The photosensors could be future versions (3'' or 4'' in diameter) of the photomultiplier tubes (PMTs) employed in XENON1T (Hamamatsu R11410-21). These sensors feature a very low intrinsic radioactivity, high quantum efficiency (QE) at 178\,nm, high gain and low dark count rate at low temperatures~\cite{Baudis:2013xva,Lung:2012pi,Aprile:2015lha}. However, albeit a proven and reliable technology, PMTs are bulky, expensive and generate a significant fraction of the radioactive background in a dark matter detector, especially in terms of radiogenic nuclear recoils~\cite{Schumann:2015cpa}. Thus, several alternative light readout schemes are under consideration. In addition, new challenges may arise in scaling up the `traditional' dual-phase scheme to the multi-ton regime. To meet these potential challenges the DARWIN R\&D programme further incorporates feasibility studies of other, non-traditional, light and charge readout concepts and novel TPC configurations, as outlined in Section~\ref{sec:r_and_d}.

\section{Science channels}
\label{sec:science}

This section outlines the science capabilities of the DARWIN facility.  Due to its low energy threshold,  ultra-low ER and NR-induced background and large target volume, DARWIN will not only be sensitive to WIMP dark matter, but also to a wide variety of other rare-event searches including solar $pp$-neutrinos, supernova neutrinos, coherent neutrino scattering, axions and axion-like-particles, and neutrinoless double beta decay~\cite{Baudis:2013qla}. Section~\ref{sec::wimp} summarises the WIMP dark matter sensitivity reach, which we have investigated under various assumptions~\cite{Newstead:2013pea,Schumann:2015cpa}. The additional physics channels are described in Section~\ref{sec::other}.

\subsection{WIMP dark matter}
\label{sec::wimp}

The primary purpose of DARWIN is to investigate dark matter interactions and a large part of our activity is focused on optimising the sensitivity for WIMP dark matter. We have shown in a recent study~\cite{Schumann:2015cpa}  that one can exploit the full discovery potential of this technique with a 40\,t LXe TPC (50\,t total, and 30\,t in the fiducial target), considering all known backgrounds listed in Section~\ref{sec:backgrounds}. These include backgrounds from detector construction materials ($\gamma$-radiation, neutrons), $\beta$-decays of $^{85}$Kr (0.1\,ppt of $^{\mathrm{nat}}$Kr) and the progeny of $^{222}$Rn (0.1\,$\mu$Bq/kg) in the liquid target, two-neutrino double beta-decays (2$\nu\beta\beta$) of $^{136}$Xe, electronic recoil interactions from low energy solar neutrinos ($pp$, $^7$Be), as well as higher energy neutrino interactions with xenon nuclei in coherent neutrino-nucleus scattering (CNNS). Under these assumptions and with an exposure of 200\,t$\times$y, we find that a spin-independent WIMP sensitivity of $2.5\times10^{-49}$\,cm$^2$ can be reached at a WIMP mass $m_\chi=40$\,GeV/$c^2$, as shown in Figure~\ref{fig::limits}~(left) on page~\pageref{fig::limits}. Increasing the exposure to 500\,t$\times$y improves this sensitivity to $\sim$$1.5\times10^{-49}$\,cm$^2$, under identical assumptions. 

Natural xenon includes two isotopes with non-zero total nuclear angular momentum, $^{129}$Xe and $^{131}$Xe, at a combined abundance of $\sim$50\%. For spin-dependent WIMP-neutron couplings and WIMP masses up to $\sim$1\,TeV/$c^2$, the searches that can be conducted by DARWIN will be complementary to those of the future high-luminosity LHC, at 14\,TeV center-of-mass energy~\cite{Malik:2014ggr}, as shown in Figure~\ref{fig::limits} (right). If the WIMP-nucleus interaction is indeed spin-dependent, DARWIN can also probe inelastic scattering, where the $^{129}$Xe and $^{131}$Xe nuclei are excited into low-lying states, with subsequent prompt deexcitation~\cite{Baudis:2013bba}.

The projected sensitivity critically depends on the ability to discriminate NR signals from ER background, as the background from low-energetic solar neutrinos cannot be mitigated by other methods. Our study assumes an ER rejection level of 99.98\% at 30\% nuclear recoil acceptance, which is a factor~5 above the one of XENON100~\cite{Aprile:2012nq} and has already been achieved by ZEPLIN-III~\cite{Lebedenko:2008gb}. Crucial requirements for reaching this rejection level include a uniform and high light yield for~S1 and an S2~signal detection with uniform electron extraction and gas amplification. The statistical fluctuations in the S1 signal close to threshold significantly affect the width of the electronic and nuclear recoil distributions. Uniformity in~S1 and S2~signal detection minimises any instrument-related fluctuations affecting the width of the S2/S1 distributions and hence the ER rejection power. While an increased light yield will also reduce the energy threshold, the dominating CNNS background will render thresholds below 5\,keV nuclear recoil energy (5\,keV$_{\textnormal{nr}}$) less relevant for the WIMP search at spin-independent cross sections below $\sim$10$^{-45}$cm$^2$. A further consideration is that the steeply falling CNNS spectrum requires the best possible energy resolution also at threshold. An energy scale derived from the charge signal or from a combination of light and charge is therefore necessary to optimise the sensitivity, as discussed in Section~\ref{sec::resolution}.

\begin{figure}[b!]
\begin{minipage}[]{0.5\textwidth}
\begin{center}
\includegraphics[width=1\textwidth]{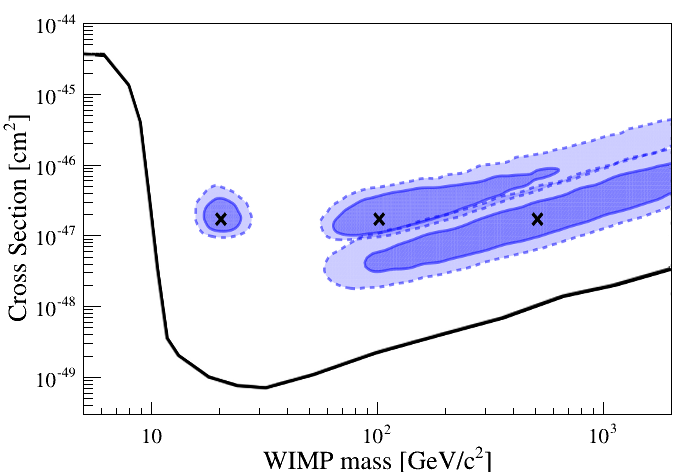}  \end{center}
\end{minipage}
\hfill
\begin{minipage}[]{0.5\textwidth}
\begin{center}
\includegraphics[width=1\textwidth]{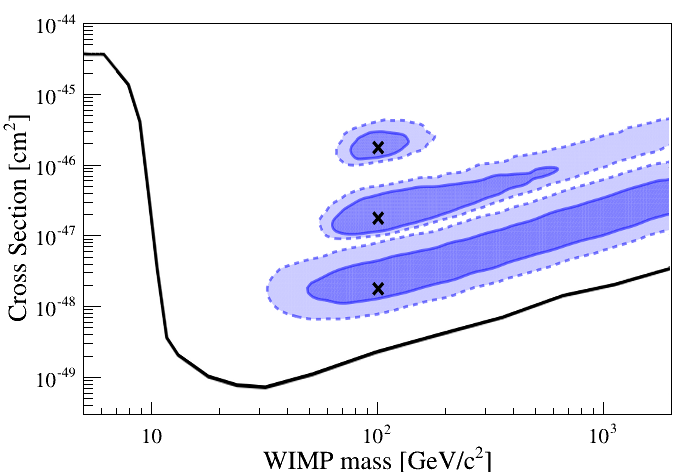} \end{center}
\end{minipage}
{\small \caption{The 1\,$\sigma$ and 2\,$\sigma$ credible regions of the marginal posterior probabilities for simulations of WIMP signals assuming various masses and spin-independent (scalar) cross sections with DARWIN's LXe target. The width and length of these contours demonstrate how well the WIMP parameters can be reconstructed in DARWIN after a 200\,t$\times$y exposure. The `$\times$' indicate the simulated benchmark models. {\bf (left)} Reconstruction for three different WIMP masses of 20\,GeV/$c^2$, 100\,GeV/$c^2$ and 500\,GeV/$c^2$ and a cross section of $2\times10^{-47}$\,cm$^2$, close to the sensitivity limit of XENON1T. {\bf (right)}  Reconstruction for cross sections of $2\times10^{-46}$\,cm$^2$, $2\times10^{-47}$\,cm$^2$ and $2\times10^{-48}$\,cm$^2$ for a WIMP mass of 100\,GeV/$c^2$. The black curve indicates where the WIMP sensitivity will start to be limited by neutrino-nucleus coherent scattering.  Figure adapted from~\cite{Newstead:2013pea}.} \label{fig:sens}}
\end{figure}

We have studied the reconstruction of WIMP properties, namely mass and scattering cross section, from the measured recoil spectra. In a numerical model, we have incorporated realistic detector parameters, backgrounds and astrophysical uncertainties~\cite{Newstead:2013pea}. Our primary study was directed towards spin-independent WIMP-nucleon interactions; however, given DARWIN's excellent sensitivity to spin-dependent interactions, especially for  $^{129}$Xe~\cite{Menendez:2012tm}, it can be extended to axial vector couplings as well. Figure~\ref{fig:sens}~(left) shows the reconstructed parameters for three hypothetical particle masses and a fixed cross section of $2 \times10^{-47}$\,cm$^2$, assuming an exposures of 200\,t$\times$y~\cite{Newstead:2013pea}. The corresponding number of events are 154, 224 and 60, for WIMP masses of 20\,GeV/$c^2$, 100\,GeV/$c^2$ and 500\,GeV/$c^2$, respectively. Using the same exposure, Figure~\ref{fig:sens}~(right) shows the reconstructed mass and cross section values for a 100\,GeV/$c^2$ WIMP and several cross sections. The study applies a conservative nuclear recoil energy threshold of 6.6\,keV$_{\textnormal{nr}}$, a Helm nuclear form factor~\cite{Vietze:2014vsa} and a Maxwell-Boltzmann distribution for the WIMP speed. The uncertainties on the dark matter halo parameters $\rho_0$=(0.3$\pm$0.1)\,GeV\,cm$^{-3}$, $v_0$=(220$\pm$20)\,km\,s$^{-1}$ and $v_{\rm{esc}}$=(544$\pm$40)\,km\,s$^{-1}$ have been marginalised over, and lead to extended regions in the mass-cross section parameter space~\cite{Strigari:2009zb,Pato:2012fw}. The parameter reconstructions were performed on a representative data set, where the number of observed events was equal to the expected number of events for the given WIMP and detector parameters. A real experiment would be subject to realisation noise, which would induce a shift in the reconstructed regions from the underlying ground truth, as quantified in~\cite{Strege:2012kv}.

The tightest constraints are obtained for WIMP masses up to a few hundred GeV/$c^2$. We also find that for masses $\geq$500\,GeV/$c^2$ only lower limits on the WIMP mass can be derived due to the fact that the shape of the nuclear recoil spectra depends on the WIMP-nucleus reduced mass.  Figure~\ref{fig:sens}  shows that, even with a large exposure such as 200\,t$\times$y, a substantial uncertainty on the reconstruction of the WIMP properties remains, depending on the mass and cross section.  The extraction of dark matter properties is complicated by the astrophysical uncertainties, in particular from the underlying phase space distribution in our Galactic halo and the local normalisation -- both of which can induce systematic errors if not properly accounted for. This systematic bias can be converted into a more manageable statistical error by introducing a parametric astrophysical model and marginalising over its parameters~\cite{Strigari:2009zb}.

\subsection{Other rare event searches}
\label{sec::other}

In this section we describe several other searches for rare events which can be pursued by multi-ton liquid xenon experiments. Due to the expected low background of electronic recoils, DARWIN will be sensitive not only to WIMPs, but also to some additional, hypothetical particles which are expected to have non-vanishing couplings to electrons. It will also be able to detect solar neutrinos, which constitute part of the background for the WIMP search channel, and neutrinos from supernova explosions in the galaxy or the Magellanic Clouds.

\subsubsection{Axions and axion-like particles}
\label{sec:axions}

Galactic axions and axion-like particles (ALPs) are well-motivated dark matter candidates~\cite{Ringwald:2012hr,Baudis:2016qwx}. Even if axions do not represent the majority of the dark matter in our Universe, they could still exist and be abundantly produced in the Sun.   By exploiting the axio-electric effect~\cite{Pospelov:2008jk,Derevianko:2010kz}, DARWIN can search for galactic and solar axions.  In this process, axions couple to the electrons of the xenon atoms in the target and lead to atomic ionisation. They can thus be detected in the electronic recoil channel, down to energies of a few keV.  The process is analogous to that of the photoelectric effect, and the expected signature would be a mono-energetic peak at the axion mass, spread only by the energy resolution of the detector.

\begin{figure}[h!]
\begin{minipage}[]{0.5\textwidth}
\begin{center}
\includegraphics*[width=1.\textwidth]{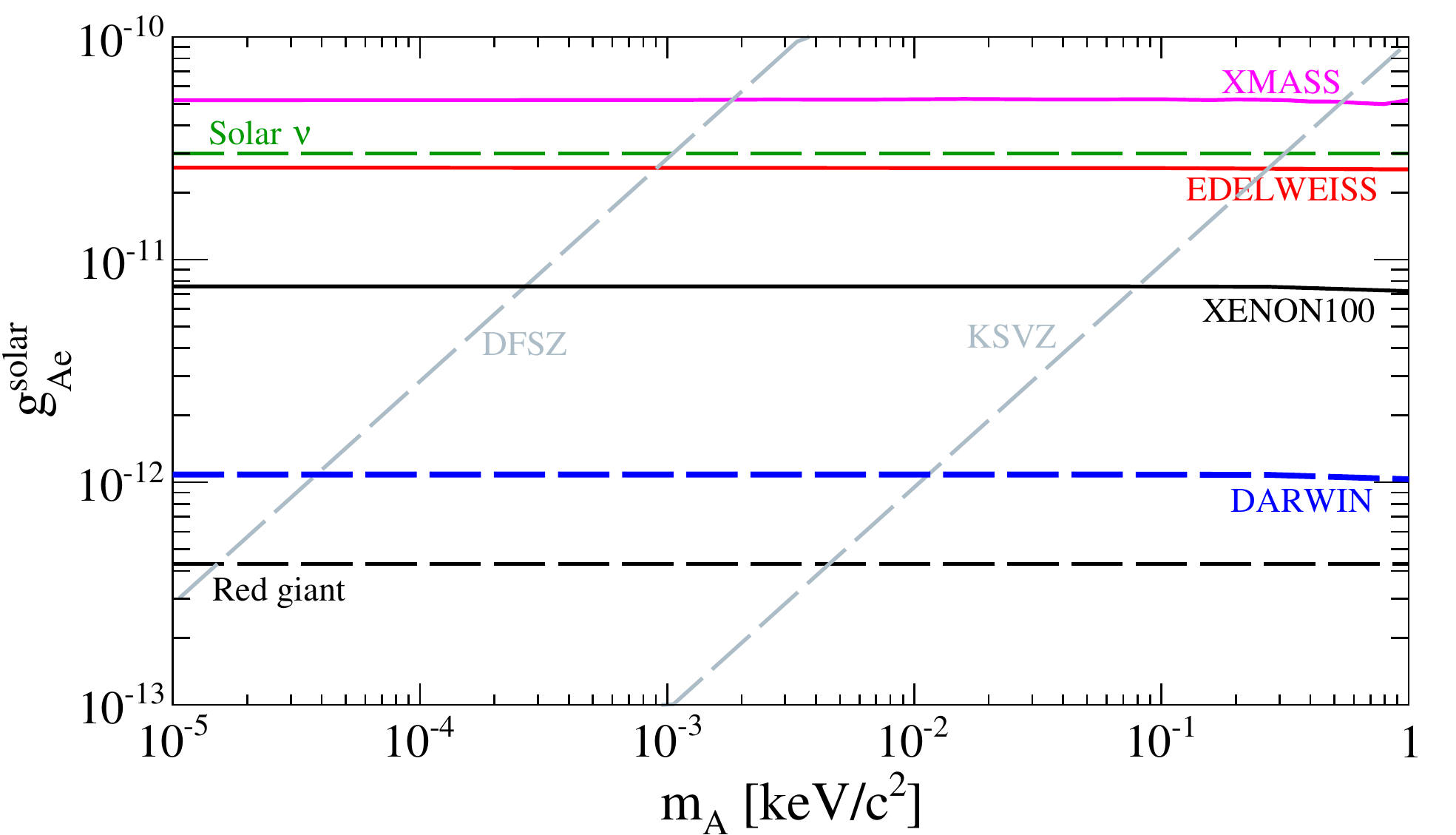}  
\end{center}
\end{minipage}
\hfill 
\begin{minipage}[]{0.5\textwidth}
\includegraphics*[width=1.\textwidth]{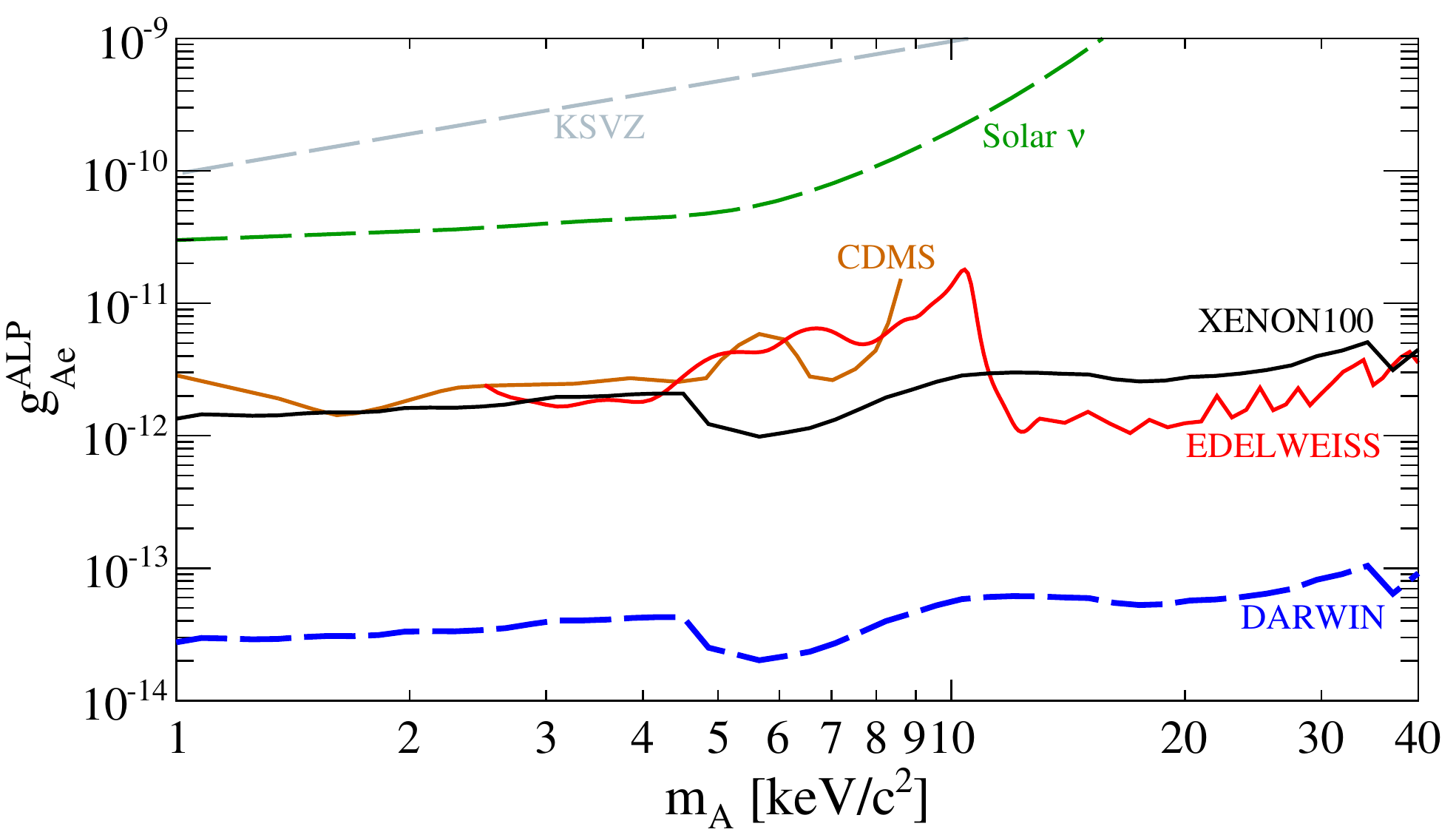}  
\end{minipage}
{\small \caption{Sensitivity of DARWIN to solar axions  {\bf (left)}  and  axion-like-particles (ALPs) {\bf (right)} which could constitute the entire galactic dark matter. While the increase in sensitivity compared to the XENON100 result~\cite{Aprile:2014eoa} is only moderate for solar axions, due to the very weak dependence on the exposure ($x^{-1/8}$), DARWIN could improve the sensitivity to galactic ALPs by almost two order of magnitude in a 200\,t\,$\times$\,y exposure. Direct upper limits from the dark matter experiments XMASS~\cite{Abe:2012ut}, EDELWEISS~\cite{Armengaud:2013rta} and CDMS~\cite{Ahmed:2009ht}, indirect limits from solar neutrinos and red giants, as well as two generic axion models, DFSZ \cite{Dine:1981rt} and KSVZ \cite{Shifman:1979if}, are also shown.
\label{fig::axions} }}
\end{figure}

XENON100 was the first to report results on solar axions and galactic ALPs using a dual-phase xenon detector~\cite{Aprile:2014eoa}. It excluded axion-electron couplings $g^{\rm{ALP}}_{Ae} \gtrsim 2\times10^{-12}$ (90\% CL) for galactic ALPs in a mass range of $1\,<\,m_A\,<40$\,keV/$c^2$. In the case of solar axions, it excluded  axion-electron couplings $g^{\rm{solar}}_{Ae} > 7.7\times10^{-12}$ (90\% CL)  in a mass range of $10^{-5}\,<\,m_A\,<1$\,keV/$c^2$, as shown in Figure~\ref{fig::axions}. 

The sensitivity of large liquid xenon detectors to axion signals was first studied in~\cite{Arisaka:2012pb}. Extrapolating from these results, and assuming the total ER background estimated in~\cite{Schumann:2015cpa}, a similar energy threshold, a 30\% superior energy resolution to XENON100, and an exposure of 200\,t$\times$y, we find that DARWIN could improve the sensitivity of XENON100 for galactic ALPs by almost two orders of magnitude, as shown in Figure~\ref{fig::axions}, right. For solar axions, the sensitivity improvement  will be more modest, equaling about one order of magnitude, see Figure~\ref{fig::axions}, left. This is due to the rather weak dependence of the coupling on the exposure (target mass $M$\,$\times$ time $T$), with $g^{\rm{ALP}}_{Ae} \propto (M\,T)^{-1/4}$ for galactic ALPs and $g^{\rm{solar}}_{Ae} \propto (M\,T)^{-1/8}$ for solar axions. The dominating background for these searches will come from irreducible solar neutrino interactions and from the $2\nu\beta\beta$ of $^{136}$Xe, see Sections~\ref{sec:solar} and~\ref{sec:bbdecay}, respectively.

\subsubsection{Solar neutrinos}
\label{sec:solar}

The most restrictive background for DARWIN's dark matter physics program will come from solar neutrino interactions, see Section~\ref{sec:nu_bg}. On the other hand, the DARWIN detector can also study neutrinos, and this capability opens up another relevant physics channel, as detailed in~\cite{Baudis:2013qla}. A precise measurement of the {\it pp}-neutrino flux would test the main energy production mechanism in the Sun, since the {\it pp}- and $^7$Be-neutrinos together account for more than 98\% of the total neutrino flux predicted by the Standard Solar Model. A total $^7$Be-neutrino flux of  $(4.84\pm0.24)\times10^9$\,cm$^{-2}$s$^{-1}$ has been measured by the Borexino experiment~\cite{Bellini:2011rx}, assuming MSW-LMA solar oscillations, a flux which was confirmed by KamLAND~\cite{Gando:2014wjd}. However, the most robust prediction of the Standard Solar Model is for the {\it pp}-neutrino flux, which is heavily constrained by the solar luminosity in photons. A high-precision real-time comparison between the solar photon luminosity and the luminosity inferred by the direct measurement of the solar {\it pp}-neutrino flux would therefore severely limit any other energy production mechanism, besides nuclear fusion, in the Sun. Borexino has recently reported the first direct measurement of the {\it pp}-flux, $(6.6\pm0.7)\times 10^{10}$\,cm$^{-2}$s$^{-1}$, with $\sim$10\% precision~\cite{Bellini:2014uqa}. 

\begin{figure}[h!]
\begin{minipage}[]{0.47\textwidth}
\begin{center}
\includegraphics*[width=1.\textwidth]{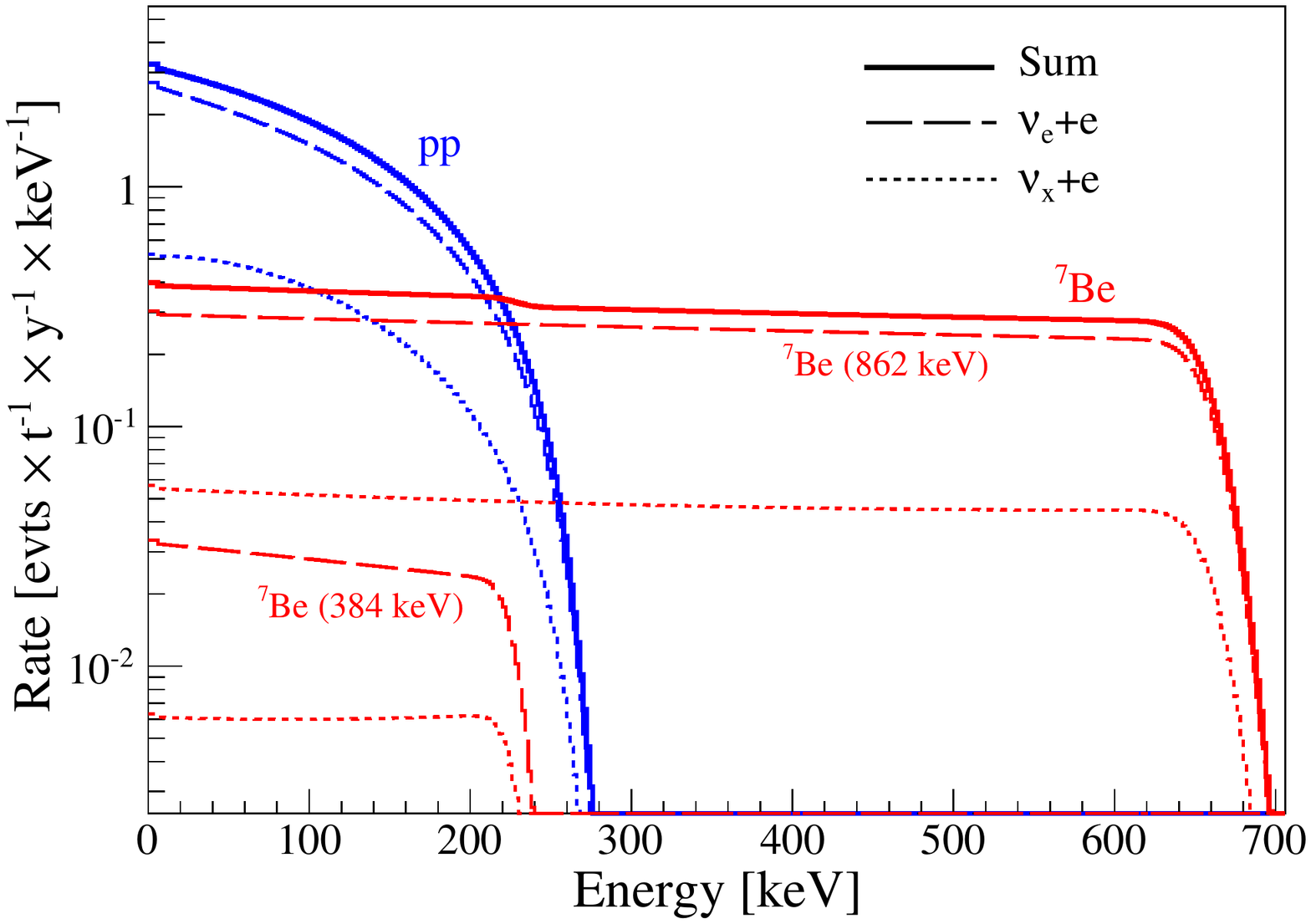}  
\end{center}
\end{minipage}
\hfill 
\begin{minipage}[]{0.52\textwidth}
\includegraphics*[width=1.\textwidth]{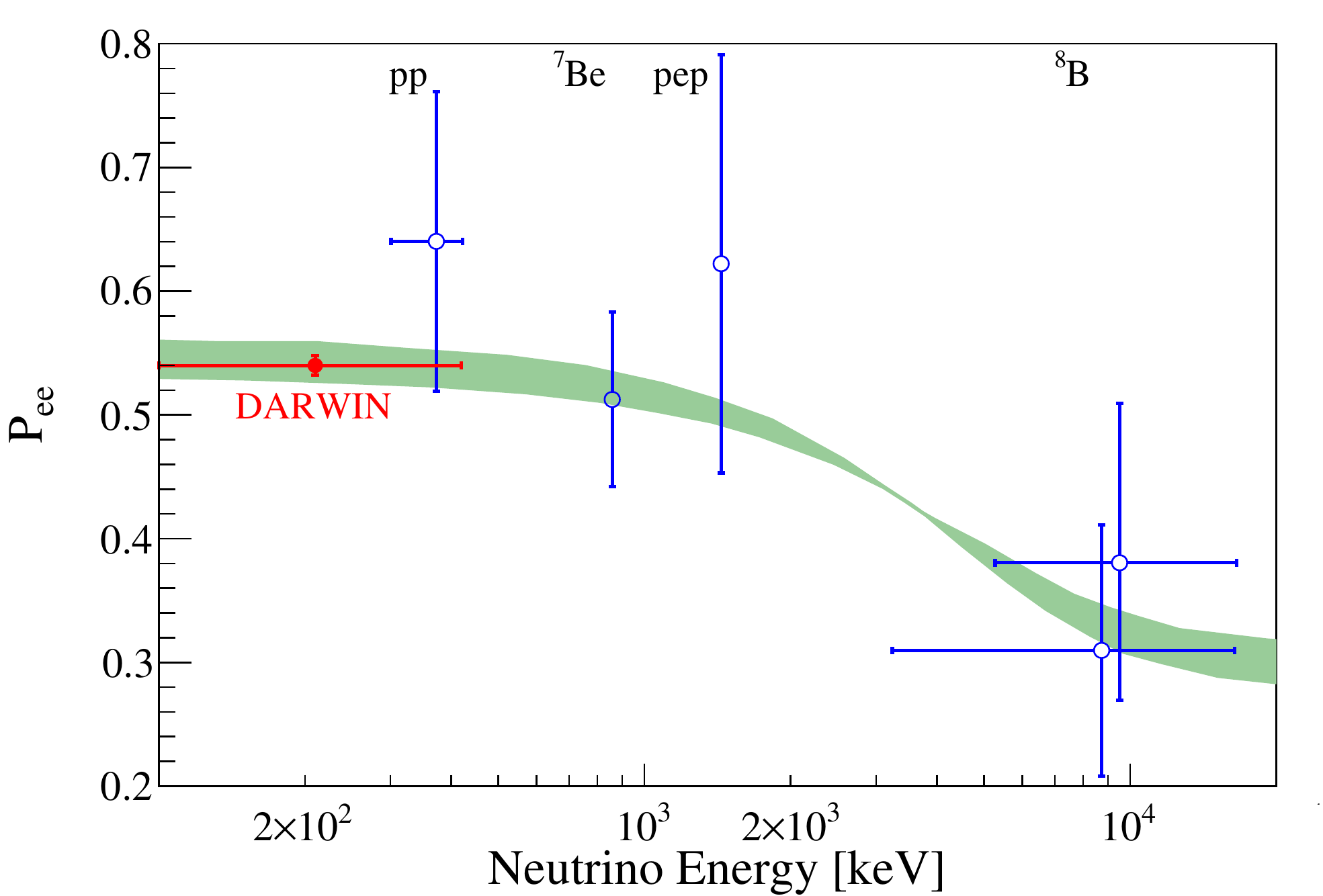}  
\end{minipage}
{\small \caption{{\bf (left)} Differential electron recoil spectra for {\it pp}- (blue) and $^7$Be neutrinos (red) in liquid xenon. The sum contribution (solid line) is split into the contributions from $\nu_e$ (dashed) and the other flavours (dotted). Figure adapted from~\cite{Baudis:2013qla}. {\bf (right)} Survival probability of solar, electron-neutrinos.  The expected sensitivity of DARWIN (red) is shown together with existing measurements from Borexino and the MSW neutrino oscillation prediction ($\pm$1$\sigma$, green) for the large mixing angle scenario~\cite{Bellini:2014uqa}. The precise measurement of the $pp$-flux with sub-percent precision with DARWIN will allow for testing neutrino and solar models.
\label{fig::solar_neutrinos} }}
\end{figure}

The detection of low-energy solar neutrinos is through elastic neutrino-electron scattering $\nu + e^- \rightarrow \nu + e^{-}$. We estimated the potential of the DARWIN detector to measure their spectrum in real time~\cite{Baudis:2013qla}. Figure~\ref{fig::solar_neutrinos}~(left) shows the recoil spectrum from {\it pp} and $^{7}$Be neutrinos. (Figure~\ref{fig::wimp_neutrino}~(left) on page~\pageref{fig::wimp_neutrino} focuses on the low energy region.) The total expected number of events above an energy threshold of 2\,keV$_\textnormal{ee}$ (electronic recoil equivalent) and below an upper limit of 30\,keV$_\textnormal{ee}$, imposed by the rising $2\nu\beta\beta$ spectrum of $^{136}$Xe, is $n_{pp}=7.2$\,events/day and $n_{^7\rm{Be}}=0.9$\,events/day. These numbers assume a fiducial target mass of 30\,tons of natural xenon and take into account the most recent values for the neutrino mixing angles~\cite{Agashe:2014kda}. More than 2$\times$10$^3$ {\it pp}-neutrino events will be observed per year,  allowing for a measurement of the flux with 2\%~statistical precision. A precision below~1\% would be reached after 5~years of data taking. DARWIN would therefore address one of the remaining experimental challenges in the field of solar neutrinos, namely the comparison of the Sun's neutrino and electromagnetic luminosities with a precision of $<$1\%~\cite{Robertson:2012ib}. The high statistics measurement of the {\it pp}-neutrino flux would open the possibility to test the solar model and neutrino properties, see Figure~\ref{fig::solar_neutrinos}~(right).  For example, non-standard neutrino interactions~\cite{Friedland:2004pp,Maltoni:2015kca} can modify the survival probability of electron neutrinos in the transition region around 1\,MeV but also at {\it pp}-neutrino energies.

\subsubsection{Neutrinoless double-beta decay}
\label{sec:bbdecay}

The question about whether neutrinos are Majorana fermions (i.e., their own antiparticles) is of intense scientific interest~\cite{Bilenky:2012qi}. The most practical investigation of the Majorana nature of neutrinos, and of lepton number violation, is through the search for neutrinoless double-beta decay ($0\nu\beta\beta$). $^{136}$Xe is an interesting $0\nu\beta\beta$-decay candidate and has an abundance of 8.9\% in natural xenon. Its $Q_{\beta\beta}$-value is at 2.458\,MeV, well above the energy-range expected from a WIMP recoil signal. 

\begin{figure}[t!]
\begin{minipage}[]{0.58\textwidth}
\begin{center}
\includegraphics*[width=1.\textwidth]{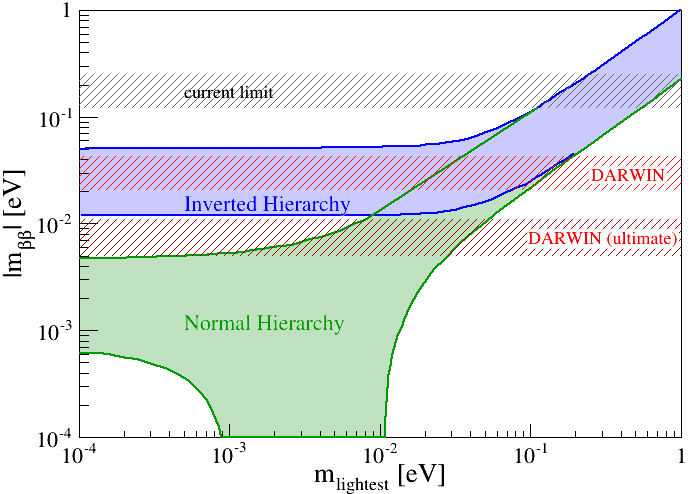}  
\end{center}
\end{minipage}
\hfill 
\begin{minipage}[]{0.40\textwidth}
\vspace{-0.4cm}
{\small \caption{Expected sensitivity for the effective Majorana neutrino mass. The sensitivity band widths reflect the uncertainties in the nuclear matrix element of the $^{136}$Xe 0$\nu\beta\beta$-decay. The `DARWIN' sensitivity assumes a 30\,t$\times$y exposure of natural xenon and a background dominated by $\gamma$-rays from detector materials. The 'ultimate' case with 140\,t$\times$y exposure assumes this background being absent, thus only $^{222}$Rn, $2\nu\beta\beta$ and $^8$B solar neutrinos contribute. For details, see~\cite{Baudis:2013qla}.  Also shown are the expected regions for the two neutrino mass hierarchy scenarios.
\label{fig::betabeta} }}
\end{minipage}
\end{figure}

Two experiments, EXO-200~\cite{Albert:2014awa} and KamLAND-Zen~\cite{Gando:2012zm,Gando:2016cqd}, have already reported very competitive lower limits on the half-life using a few hundred kilograms of $^{136}$Xe. Even without isotopic enrichment, DARWIN's target contains more than 3.5\,t of $^{136}$Xe and can be used to perform a search for its $0\nu\beta\beta$-decay in an ultra-low background environment. The main challenge for this measurement will be to optimise the detector's sensors and readout electronics to perform at both the $\mathcal{O}(10)$\,keV energy-scale and at the $\mathcal{O}(1)$\,MeV scale relevant for the expected 0$\nu\beta\beta$-decay peak.  Once a resolution $\sigma/E$\,$\sim$\,1-2\% at $\sim$2.5\,MeV   is achieved and the background is reduced by a strong fiducialisation or the selection of ultra-low radioactivity detector materials, DARWIN's $0\nu\beta\beta$ sensitivity will become comparable to other future ton-scale experiments. In Figure~\ref{fig::betabeta}, we show its reach for the effective Majorana neutrino mass~$|m_{\beta\beta}|$ versus the mass of the lightest neutrino, as calculated in~\cite{Baudis:2013qla}. The corresponding sensitivities to the half-life of the decay are \mbox{T$_{1/2}^{0\nu}>$~5.6$\times$10$^{26}$\,y} and \mbox{T$_{1/2}^{0\nu}>$~8.5$\times$10$^{27}$\,y} (at 90\% C.L.) for assumed natural xenon exposures of 30\,t$\times$y and 140\,t$\times$y, respectively. The latter value assumes that the material backgrounds can be completely mitigated and that the only background sources are 0.1\,$\mu$Bq/kg of $^{222}$Rn, $2\nu\beta\beta$-decays and interactions of solar $^8$B neutrinos. The width of the bands reflect the uncertainties in the nuclear matrix element calculations of the $^{136}$Xe 0$\nu\beta\beta$-decay. We expect smaller exposures compared to the WIMP search, caused by a more stringent fiducialisation to reach the required background level at $Q_{\beta\beta}$. Other rare nuclear processes, such as the double-beta decays of $^{134}$Xe, $^{126}$Xe and $^{124}$Xe, might be observable as well~\cite{Barros:2014exa}. 

\subsubsection{Coherent neutrino-nucleus scattering}
\label{sec:cnns}

The rate of low-energy signals in all multi-ton WIMP detectors will eventually be dominated by interactions of cosmic neutrinos via coherent neutrino-nucleus scattering (CNNS)~\cite{Strigari:2009bq}. DARWIN will be able to detect and study this yet-unobserved standard model process, which produces a nuclear recoil signal like the WIMP. (For the implications on the WIMP search, see Section~\ref{sec:nu_bg}.) The largest CNNS rate comes from the relatively high-energy $^8$B solar neutrinos which produce nuclear recoils $\leq$3\,keV$_\textnormal{nr}$. Neutrinos from the solar hep-process induce a similar maximal recoil energy but their flux is much lower. Events from neutrinos created in the upper atmosphere and from the diffuse supernova neutrino background will extend to slightly higher energies of $\cal O$(10)\,keV$_\textnormal{nr}$, however at significantly lower rates.

Because LXe detectors operate at low energy thresholds, even the low-energy CNNS signal from $^8$B neutrinos is readily accessible. LUX has demonstrated an energy threshold of 1.1\,keV$_\textnormal{nr}$ in a re-analysis of their data~\cite{Akerib:2015rjg}, and XENON10~reached a threshold of 1.4\,keV$_\textnormal{nr}$ using an energy scale based on the S2~signal only~\cite{Angle:2011th}.  Such thresholds would lead to an  observed rate of $\sim$90 events\,t$^{-1}$y$^{-1}$  from $^8$B neutrinos. The rate from atmospheric neutrinos will be much lower, around 3$\times$10$^{-3}$ events\,t$^{-1}$y$^{-1}$~\cite{Strigari:2009bq}.  Being a standard model process, the cross section of coherent neutrino-nucleus scattering can be calculated to a high precision~\cite{Drukier:1983gj,Cabrera:1984rr}. A deviation of the experimentally measured value from the expectation is thus a clear sign of new physics, e.g., from oscillations into sterile neutrino states or from electromagnetic interactions of the neutrino~\cite{Giunti:2014ixa}.

\subsubsection{Galactic supernova neutrinos}\label{sec:sn}

Neutrinos and anti-neutrinos of all flavours are emitted by core-collapse supernovae in a burst lasting a few tens of seconds~\cite{Mirizzi:2015eza}. The measurement of their energy spectrum and time structure can provide information about supernovae properties, but also about the intrinsic properties of neutrinos~\cite{Horowitz:2003cz}. 

One avenue to detect such neutrinos is via their coherent scattering off xenon nuclei in DARWIN. In contrast to large water Cherenkov or liquid scintillator detectors, where the most significant interaction is from $\overline{\nu}_e$ via inverse $\beta$-decay on free protons, $\overline{\nu}_e + p \to e^+ + n$, DARWIN would be sensitive to all six neutrino species $\nu_x$ via neutral current interactions. The measured recoil spectrum would provide direct information about the neutrino energy spectrum, which is important to study neutrino oscillations etc.~\cite{Horowitz:2003cz}. The expected nuclear recoil event rate in a LXe detector depends on the distance of the supernova explosion, the progenitor mass, the neutrino emission spectrum and on the differential neutrino-nucleus scattering cross section. It also depends on the detector properties, such as its fiducial mass, energy threshold, and detection efficiency for nuclear recoils. 

The XMASS collaboration has calculated the expected event rates in their detector~\cite{XMASS:2016cmy} and finds results consistent with those of~\cite{Chakraborty:2013zua}. However, XMASS is a single-phase detector and thus greatly limited in its sensitivity to this channel due to the higher energy threshold compared to an S2-based analysis.  For DARWIN, we expect between~10 and~20 events\,t$^{-1}$ from a supernova at a distance of 10\,kpc, depending on the supernova neutrino emission model and progenitor mass~\cite{Shayne:2016}. Due to the extremely transient nature of a neutrino burst, lasting only a few seconds, the background in the low-energy S2 range can be expected to be negligible and will allow us to use a larger fiducial target~\cite{Angle:2011th,Aprile:2016wwo}. Thus, with $\mathcal{O}(100)$~neutrino events from a Galactic supernova, DARWIN will be a supernova neutrino detector that can contribute to the Supernova Early Warning System SNEWS~\cite{Antonioli:2004zb}. By looking at the time evolution of the event rate from a nearby supernova, DARWIN could possibly distinguish between different supernova models~\cite{Shayne:2016} and provide complementary neutrino flavour-independent information to supernova physics.

\section{Expected backgrounds}\label{sec:backgrounds}

As a detector searching for rare events, DARWIN requires a very good understanding of all possible background sources and an extremely low absolute background level. The inner detector will hence be surrounded by a large, instrumented water shield, which passively reduces the environmental radioactivity as well as muon-induced neutrons, and acts as an active Cherenkov muon veto. Massive shields made from lead and copper are neither practical nor cost-effective at this scale. Typical ambient $\gamma$-ray and radiogenic neutron fluxes of 0.3\,cm$^{-2}$s$^{-1}$ and  $9\times10^{-7}$\,cm$^{-2}$s$^{-1}$~\cite{Haffke:2011fp} are reduced by a factor of 10$^6$ after 3\,m and 1\,m of water shield, respectively~\cite{Aprile:2014zvw}, and can therefore be considered negligible. The $\gamma$-induced background from the radioactivity of detector materials is also irrelevant for WIMP searches at the DARWIN mass-scale~\cite{Schumann:2015cpa}.

Here we discuss the most important background sources for DARWIN, namely the backgrounds related to cosmogenic and radiogenic neutrons, background sources intrinsic to the target material, and finally neutrino-induced backgrounds.

\subsection{Neutron backgrounds}

A major background source for the WIMP search, where the expected signature is nuclear recoils from elastic WIMP-nucleus collisions, are neutron-induced nuclear recoils. The neutrons mainly originate from $(\alpha$,$n$) reactions and spontaneous fission of heavy isotopes in detector materials, and can also be generated by interactions of cosmic ray muons that penetrate deep underground. Single-scatter nuclear recoils are in principle indistinguishable from a WIMP signal, hence the absolute neutron fluxes from all sources must be minimised {\sl a priori}, e.g., by selecting materials with low intrinsic radioactive contamination~\cite{Heusser:1995wd}. Remaining neutron backgrounds will be reduced by self-shielding and by efficient multiple-scatter rejection, thanks to the large size and the excellent position resolution of the detector. DARWIN will reduce the muon-induced nuclear recoil background by its underground location and by using an active Cherenkov muon veto. We estimate that at the depth of the LNGS laboratory (3600\,meters water equivalent) an active water shield of $\sim$14\,m diameter will reduce the rate of cosmogenic neutrons to negligible levels. Most critical are interactions from $(\alpha,n)$ and fission neutrons from $^{238}$U and $^{232}$Th decays in detector components close to the LXe, such as PTFE reflectors or photosensors.

The expected neutron background was studied for a 40\,t target mass, where the detector is mostly made of copper, PTFE, and photosensors~\cite{Schumann:2015cpa}. The neutron energy spectra and yields  in these materials were calculated taking into account their composition and reference levels of $^{238}$U and $^{232}$Th. Because secular equilibrium in the primordial decay chains is usually lost in processed materials, the  $^{238}$U and $^{232}$Th activities are often determined via mass spectrometry or neutron activation analysis, while the activities of the late part of these chains, $^{226}$Ra,  $^{228}$Ac, $^{228}$Th, are determined via gamma spectrometry using ultra-low background high-purity germanium (HPGe) detectors~\cite{Baudis:2011am,Heusser:1995wd,Heusser:2006,Heusser:2015ifa,Sivers:2016yvm}. With the detector material and radioactivity assumptions from~\cite{Schumann:2015cpa}, we obtain an expected single-scatter nuclear recoil rate of about $3.8\times10^{-5}$\,events\,t$^{-1}$\,y$^{-1}$\,keV$_\textnormal{nr}^{-1}$, in the central detector region of 30\,t. This background level is sufficiently low for an exposure of 200\,t$\times$y at $\sim$30\% nuclear recoil acceptance, but it requires us to identify materials with reduced radioactivity compared to currently measured levels. This is particularly true of the light reflector PTFE. This material contributes to the background significantly through $(\alpha,n)$ reactions due to the presence of $^{19}$F.  The neutron background could be further reduced by stronger fiducialisation, as well.

\subsection{Xenon-intrinsic backgrounds}\label{sec::intrinsic}

\begin{figure}[b!]
\begin{minipage}[]{0.5\textwidth}
\begin{center}
\includegraphics*[width=1.\textwidth]{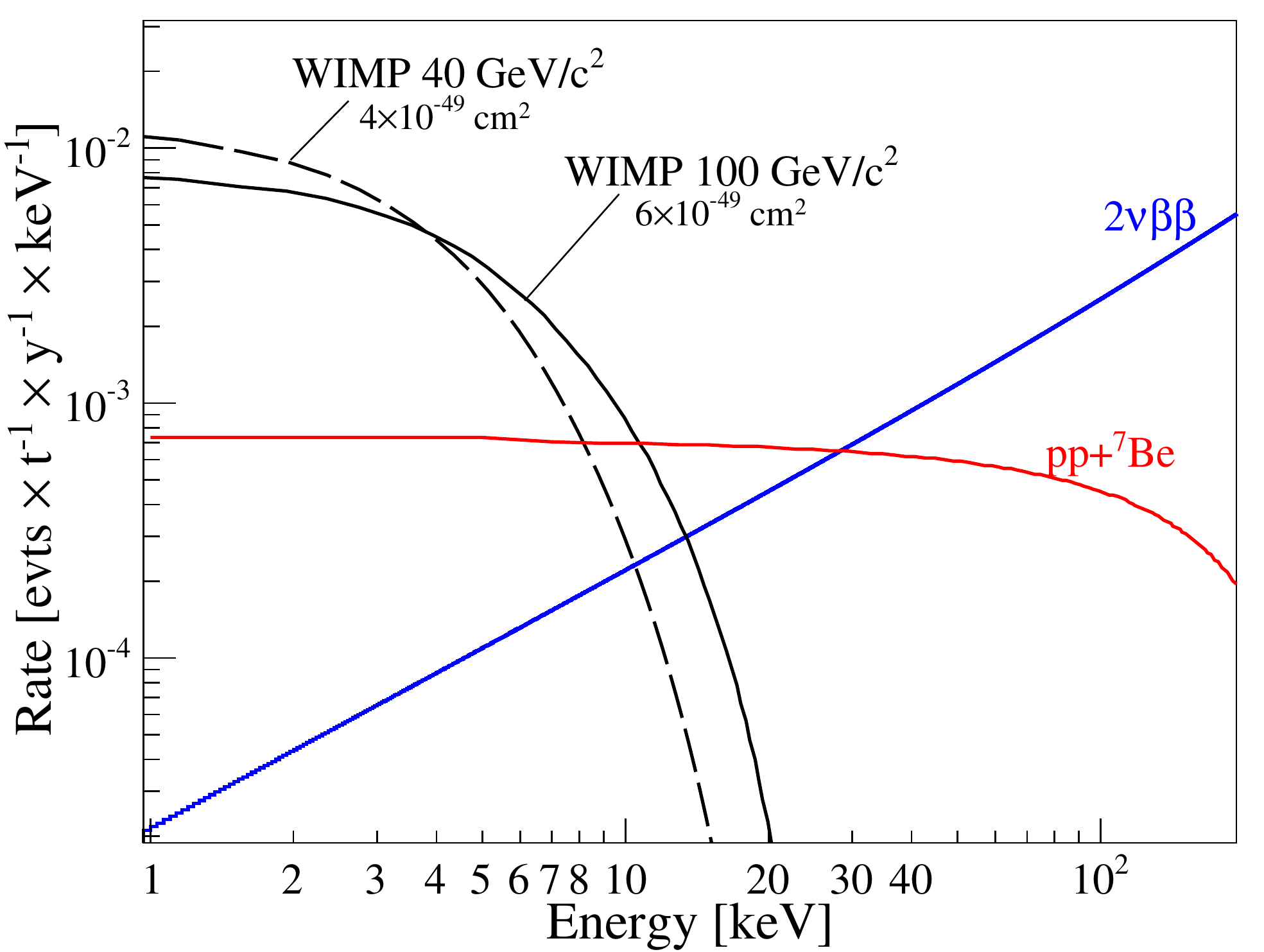}  
\end{center}
\end{minipage}
\hfill
\begin{minipage}[]{0.5\textwidth}
\begin{center}
\includegraphics*[width=1.\textwidth]{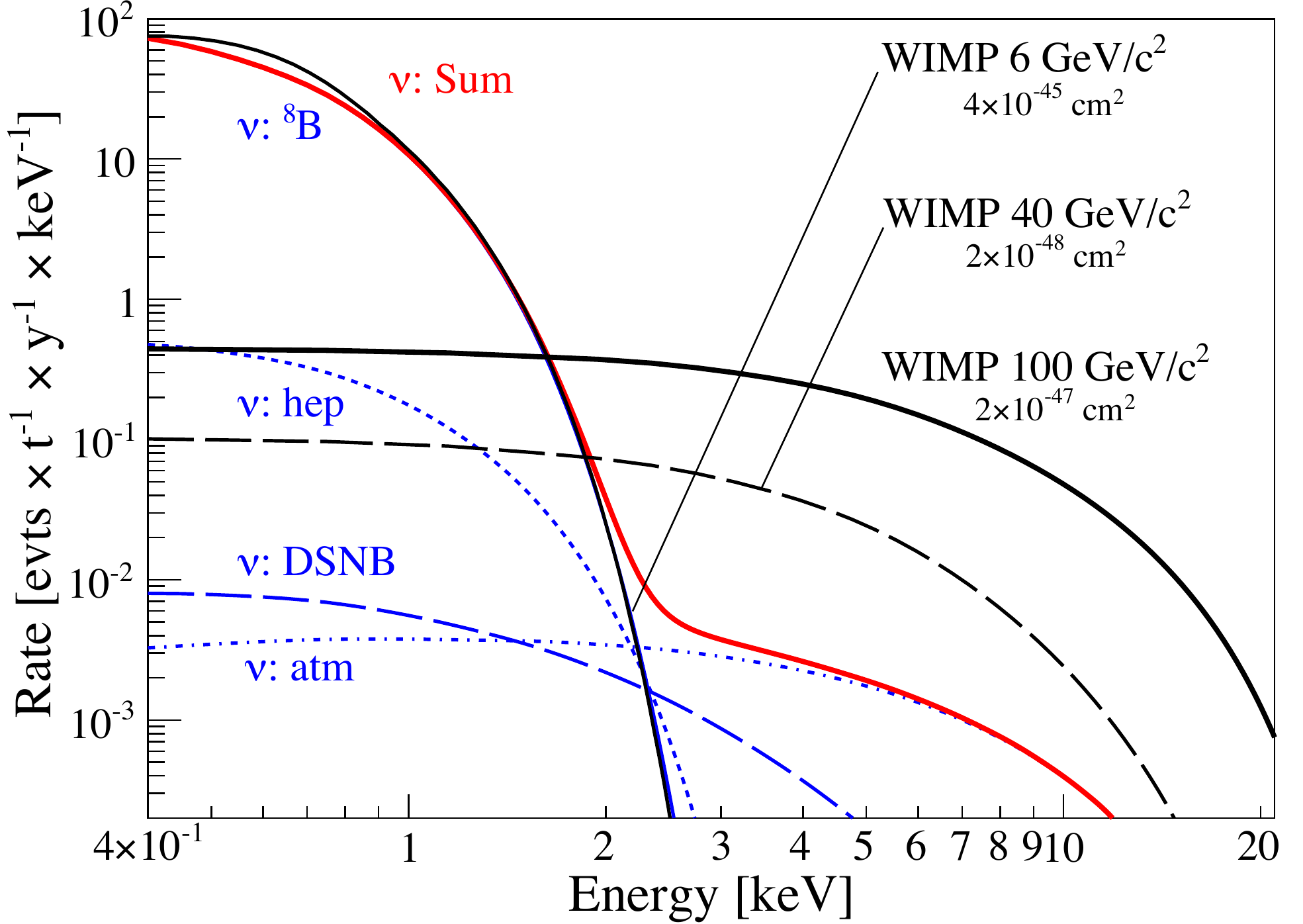}  
\end{center}
\end{minipage}
{\small \caption{{\bf (left)}  Summed differential energy spectrum for $pp$ and  $^{7}$Be neutrinos (red) undergoing neutrino-electron scattering in a LXe detector. Also shown are the electron recoil spectrum from the double-beta decay of $^{136}$Xe (blue), as well as the expected nuclear recoil spectrum from WIMPs for a spin-independent WIMP-nucleon cross section of $6\times10^{-49}$\,cm$^2$ (solid black) and $4\times10^{-49}$\,cm$^2$ (dashed black) and  WIMP masses of 100\,GeV/c$^2$ and 40\,GeV/c$^2$, respectively.  A 99.98\% discrimination of ERs at 30\% NR acceptance is assumed and the recoil energies are derived using the S1 signal only (see~\cite{Baudis:2013qla}).  {\bf (right)} The  differential nuclear recoil spectrum from coherent scattering of neutrinos (red)  from the Sun, the diffuse supernova background (DSNB), and the atmosphere (atm), compared to the one from WIMPs  for various masses and cross sections (black). The coherent scattering rate will provide an irreducible background for low-mass WIMPs, limiting the cross section sensitivity to $\sim$$ 4\times10^{-45}$\,cm$^2$ for WIMPs of 6\,GeV/$c^2$ mass, while WIMP masses above $\sim$10\,GeV/$c^2$ will be significantly less affected. No finite energy resolution but a 50\%~NR acceptance is taken into account. Figure from~\cite{Baudis:2013qla}. \label{fig::wimp_neutrino}}}
\end{figure}

In dark matter detectors based on liquefied noble gases, radioactivity intrinsic to the WIMP target, such as $^{39}$Ar, $^{85}$Kr and $^{222}$Rn provide sources of ER backgrounds.  The activation of the xenon gas itself by exposure to cosmic rays becomes irrelevant after underground storage for a few months~\cite{Baudis:2015kqa}. As $^{39}$Ar is absent in LXe, the main challenges for DARWIN are  $^{85}$Kr  and $^{222}$Rn.  $^{85}$Kr is an anthropogenic radioactive isotope present in noble liquids extracted from air.  Currently achieved $^{\rm{nat}}$Kr-levels after purification using krypton distillation or gas chromatography are (1.0$\pm$0.2)\,ppt by XENON100~\cite{Lindemann:2013kna}, with a gas chromatography and mass spectrometry detection limit of 0.008\,ppt, (3.5$\pm$1.0)\,ppt by LUX~\cite{Akerib:2013tjd} and $<$2.7\,ppt by XMASS~\cite{Abe:2013tc}. The new $^{\rm{nat}}$Kr-removal apparatus of XENON1T has recently delivered a sample with a concentration of $<$0.03\,ppt in a test run~\cite{phd_rosendahl,master_stolzenburg}. This is a factor of~3 below the 0.1\,ppt assumed for the WIMP sensitivity study presented in Section~\ref{sec::wimp}. Even at such low $^{\rm{nat}}$Kr concentrations, the background from $2\nu\beta\beta$ decays of $^{136}$Xe in a natural xenon target only contributes at a much lower level~\cite{Schumann:2015cpa}, with a spectrum that decreases towards the threshold (see Figure~\ref{fig::wimp_neutrino}, left).

$^{222}$Rn is part of the $^{238}$U natural decay chain and constantly emitted by detector surfaces, which therefore have to be selected for low Rn-emanation~\cite{Rau:2000,Kiko:2001}. The $^{222}$Rn concentrations in LXe achieved so far are 65\,$\mu$Bq/kg in XENON100~\cite{Aprile:2012nq}, 32\,$\mu$Bq/kg in LUX~\cite{Akerib:2013tjd}, 22\,$\mu$Bq/kg in PandaX-I~\cite{Xiao:2014xyn}, 9.8\,$\mu$Bq/kg in XMASS~\cite{Abe:2013tc}, and $(3.65\pm0.37)\,\mu$Bq/kg in EXO \cite{Albert:2015nta}. With the exception of EXO, none of these experiments was particularly optimised for low radon emanation. The target concentration for XENON1T is $\sim$10\,$\mu$Bq/kg, and the smaller surface-to-volume ratio will further help with the reduction in larger detectors. Nonetheless, we anticipate that achieving a low radon level will be the largest background reduction challenge. Concentrations of $\sim$0.1\,$\mu$Bq/kg must be achieved to probe WIMP-nucleon cross sections down to a few 10$^{-49}$cm$^2$, assuming an S2/S1-based rejection of ERs at the $2 \times 10^{-4}$~level at 30\% NR acceptance~\cite{Schumann:2015cpa}. Such rejection levels have already been achieved~\cite{Akimov:2011tj}.  The emanation of $^{220}$Rn, a part of the $^{232}$Th chain, could lead to similar backgrounds. However, it has a considerably shorter half-life than $^{222}$Rn and is observed to be less abundant in existing LXe detectors, hence its impact is considered sub-dominant. 

At DARWIN's current background goal of 0.1\,ppt of $^{\rm{nat}}$Kr in Xe, 0.1\,$\mu$Bq/kg of $^{220}$Rn and a target of natural $^{136}$Xe abundance, a total Xe-intrinsic background rate of $\sim$17\,events\,t$^{-1}$\,y$^{-1}$ is expected in a 2-10\,\,keV$_\textnormal{ee}$ WIMP search energy interval~\cite{Schumann:2015cpa}. An ER rejection efficiency of $2 \times 10^{-4}$ reduces this rate to 3.5\,$\times$\,$10^{-3}$\,events\,t$^{-1}$\,y$^{-1}$.

\subsection{Neutrino backgrounds}\label{sec:nu_bg}

Neutrino signals in DARWIN will provide important science opportunities (see Sections~\ref{sec:solar}, \ref{sec:cnns} and \ref{sec:sn}). However, they will also constitute the ultimate background source for many searches, especially for dark matter, as the neutrino flux can neither be shielded nor avoided by the design of the experiment. Solar $pp$-neutrinos (and $^7$Be-neutrinos at the $\sim$10\% level) will contribute to the electronic recoil background via neutrino-electron scattering at the level of $\sim$26~events\,t$^{-1}$y$^{-1}$ in the low-energy, dark matter signal region of the detector~\cite{Baudis:2013qla}. Solar neutrinos will thus become a relevant background source at WIMP-nucleon cross sections  below 10$^{-48}$cm$^2$ and electronic recoil rejection levels around  99.98\% are required to reach the envisaged ultimate WIMP sensitivity~\cite{Schumann:2015cpa}, reducing the solar neutrino background to 5.2\,$\times$\,$10^{-3}$\,events\,t$^{-1}$\,y$^{-1}$ in a 2-10\,\,keV$_\textnormal{ee}$ WIMP search region. Figure~\ref{fig::wimp_neutrino}~(left) compares the combined background from $pp$ and $^7$Be solar neutrinos with two WIMP spectra.

Neutrino-induced nuclear recoils from coherent neutrino-nucleus scatters cannot be distinguished from a WIMP-induced signal, if no additional (statistical) discriminants such as track directionality are taken into account~\cite{Grothaus:2014hja}. The $^{8}$B solar neutrinos yield up to 10$^3$\,events\,t$^{-1}$y$^{-1}$ for heavy targets such as xenon~\cite{Strigari:2009bq}, however, all the events are at very low recoil energies below 4\,keV, see Figure~\ref{fig::wimp_neutrino}~(right).  About 90~events\,t$^{-1}$y$^{-1}$ are expected above 1\,keV$_\textnormal{nr}$. Nuclear recoils from atmospheric neutrinos and the diffuse supernovae neutrino background will yield event rates which are orders of magnitude lower but at slightly higher recoil energies. These will dominate the measured spectra at WIMP-nucleon cross sections around 10$^{-49}$\,cm$^2$ for WIMP masses above $\sim$10\,GeV/$c^2$~\cite{Strigari:2009bq,Gutlein:2010tq,Billard:2013qya,Anderson:2011bi,Baudis:2013qla,Schumann:2015cpa}. The total observed rates will strongly depend on the detector's energy threshold and energy resolution.

\section{Design considerations and associated research and development}
\label{sec:r_and_d}

DARWIN will incorporate techniques which were successfully probed in the current  generation of liquid xenon detectors and which will be tested in upcoming dual-phase dark matter experiments.  At the same time, new design features will be evaluated and possibly implemented. Some of these were demonstrated  with the XENON1T Demonstrator~\cite{ref::xe1t_demonstrator}. It established the ability to drift electrons over distances of~1\,m in LXe, high-speed purification up to 100\,standard liters per minute (slpm)~\cite{Aprile:2012jh}, and high-voltage capabilities beyond~$-$100\,kV.  Valuable input is also obtained from projects pursued outside the DARWIN consortium which employ TPCs filled with LXe to address various physics questions~\cite{Akerib:2012ys,Albert:2014awa}. In addition to these full-scale TPCs, several smaller instruments, with an R\&D program focused at specific questions, are in place at various DARWIN institutions.
The detailed design of DARWIN is not yet finalised, and further R\&D towards such an ultimate WIMP detector is needed. In the following sections, we discuss  potential designs based on state-of-the art concepts and technologies, and  introduce some non-standard concepts as well.

\subsection{Cryostat and time projection chamber}  
\label{sec::cryostat_tpc}

To reach the desired WIMP sensitivity within a reasonable timescale, DARWIN  requires a total (target) LXe mass of $\sim$50 (40)\,tons, and hence a cylindrical detector with linear dimensions $>$2.5\,m~\cite{Schumann:2015cpa}. The WIMP search target after fiducialisation would be around 30\,tons. The vacuum-insulated cryostats must be constructed from materials with a very low specific radioactivity level, where the cleanest available metal is copper. However, it imposes tight mechanical constraints, and, given the excellent self-shielding capabilities of LXe, it might be beneficial to use alternatives such as titanium or stainless steel instead, as in the XENON, LUX/LZ and PandaX projects. 

\begin{figure}[t!]
\begin{minipage}[t]{0.55\textwidth}
\begin{center}
\includegraphics*[width=1.\textwidth]{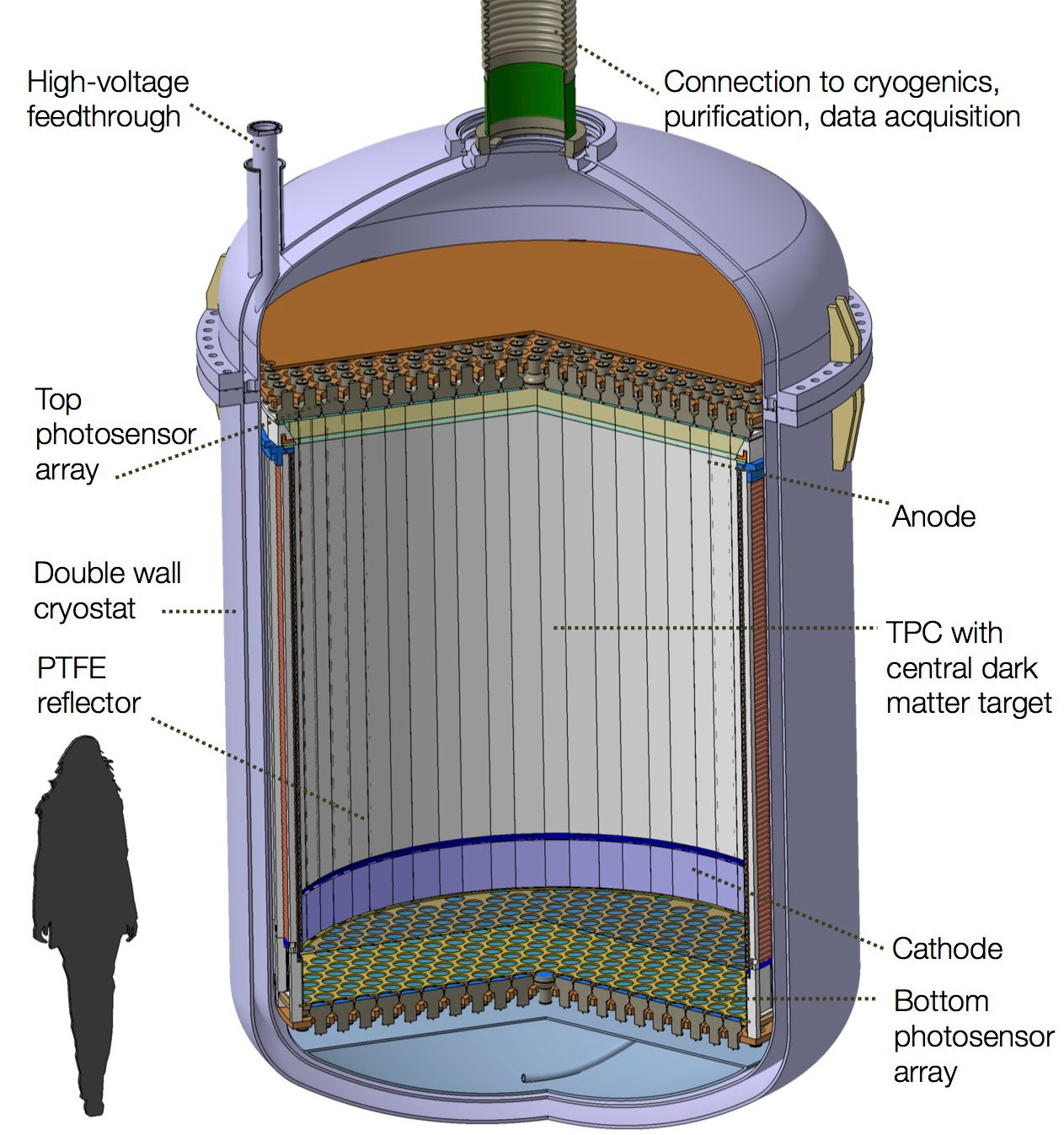}  
\end{center}
\end{minipage}
\hfill
\begin{minipage}[b]{0.4\textwidth}
\caption{A possible realisation of a $\sim$50\,t (40\,t) total (target) LXe mass DARWIN detector, inside a double-walled stainless steel cryostat. The TPC is surrounded by highly reflective PTFE walls, closed by the cathode and anode electrodes on bottom and top, respectively. The sketch shows a TPC with two photosensor arrays made of circular PMTs with 3''\,diameter. The final sensor type, however, is not yet defined and all details regarding the cryostat and TPC are subject to R\&D.   \label{fig::tpc} }
\end{minipage}
\end{figure}

All dual-phase LXe TPCs in current dark matter projects utilise the same concepts and mostly differ from each other in the details (high-voltage generation, aspect ratio, PMT granularity, liquid level control, etc.). DARWIN, in its baseline configuration, will feature this well-established dual-phase TPC design scheme with light detected by photosensor arrays above and below the LXe target, see Figure~\ref{fig::tpc}. The light collection efficiency is constant for a fixed height-diameter ratio. With an optimal design of the reflecting inner TPC surfaces, it is only affected by the LXe absorption length. The working hypothesis of DARWIN's baseline design is that the absorption length can be kept much larger than the TPC diameter by continuous purification of the xenon, see Section~\ref{sec::cryogenics}. Under this assumption, with state-of-the-art PMTs, it is expected that the currently achieved thresholds of $\sim$1\,keV$_\textnormal{nr}$ \cite{Akerib:2015rjg} can also be established with DARWIN. To cope with the possibility of smaller values for the absorption length -- or, alternatively, to further increase the light collection efficiency -- a potential scheme with the TPC surrounded by photosensors in $\sim$4$\pi$, similar to a single-phase detector, is being evaluated as part of the DARWIN R\&D program. This option is outlined in Section~\ref{sec:light}, which also discusses alternative photosensor technologies. A novel scheme relying on the concept of liquid hole multipliers (LHMs), with a potentially significant light yield improvement, is discussed in Section~\ref{sec:light}, as well.  

Insulating materials are essential to construct the TPC, as components biased with very high voltages above $-$100\,kV (cathode, field shaping electrodes) must be supported and insulated from grounded components. The primary choice  is PTFE providing excellent insulation, good UV reflectivity~\cite{ref::ptfe_reflectivity}, reasonable mechanical strength, and low radioactivity.  A possible cylindrical DARWIN TPC of 260\,cm diameter and height, enclosing a target mass of 40\,t of xenon, is illustrated in Figure~\ref{fig::tpc}.

The type and dimension of the light sensors installed on the two arrays, above and below the target  are still an active part of the DARWIN study. Under the assumption that the charge signal is detected via proportional scintillation in the gas phase, the 40\,t LXe TPC would require $\sim$1800~sensors of 3''~diameter ($\sim$1000 of~4'') assuming the use of identical, circular photosensors on both arrays. If available,  larger low-radioactivity photosensors on the bottom array could reduce the number of channels, as discussed in Section~\ref{sec:light}. A finite detector granularity on the top array is required for the $xy$-vertex identification.

\subsection{High voltage system} 
\label{sec::hv}

To create an electron drift field of $\sim$0.5\,kV/cm across the TPC, the cathode must be biased by a high, negative voltage. The field homogeneity is ensured by a set of circular field shaping rings, interconnected with high-ohmic resistors, realising a voltage divider which is gradually approaching ground potential. For a TPC of  2.6\,m length, a cathode potential of $130$\,kV is required to establish the design field. We note that the TPC can  be successfully operated at lower fields, as well (LUX: 0.18\,kV/cm~\cite{Akerib:2013tjd}); this reduces the field quenching, resulting in larger S1 signals. On the other hand, the lower electron drift velocity increases the pile-up rate during calibration runs and there are indications that the S2/S1~discrimination power deteriorates for reduced drift fields~\cite{Lebedenko:2008gb}.

While the field shaping rings, which are made from massive copper with a smooth surface, do not impose a problem for the high voltage, the cathode electrode and the HV-feedthrough must withstand the high operation potentials. In order to optimise the optical transparency, the cathode  will be made of single wires of $\sim$100\,$\mu$m diameter, spot-welded to a sturdy, low-background metal frame. The cathode currently operated in XENON1T can be seen as a first prototype.  A low-radioactivity feedthrough with demonstrated  stable operation in LXe up to $\sim$130\,kV has been built for XENON1T.  As noted above, it would allow for reaching the design drift field of $\sim$0.5\,kV/cm, similar to the one in XENON100~\cite{Aprile:2012nq}. The extraction field across the liquid-gas interface requires a moderate, positive anode bias voltage for 100\% electron extraction efficiency~\cite{Aprile:2013blg}, which is typically between $+5$ and $+10$\,kV, depending on the distance between the gate electrode, at ground potential, and the anode. The main challenge is to keep the anode, with a diameter of $\sim$2.6\,m, parallel to the liquid surface, and to maintain a constant gap between gate and anode. This is required for achieving a homogenous S2~response across the detector's surface. The LHM-based TPC scheme discussed in Section~\ref{sec:light} may provide a non-traditional solution.

The electrostatic configuration of DARWIN's TPC field cage, which is composed of very large (grid electrodes diameter) and small (wire diameter) elements, is being designed with {\it KEMField}~\cite{Gluck:2013taa}. This simulation tool, developed within the KATRIN collaboration, has been adapted for dual-phase TPCs. Optimised for the  simulation of electric fields in large-scale geometries with small-scale structures, it takes advantage of the Boundary Element Method (BEM) to compute electric fields and potentials with the highest precision. For DARWIN-type geometries, BEM performs superior to Finite Element Methods (FEM), that require large computer memory for the meshing of the 3D-space.

\subsection{Cryogenic and purification systems}  
\label{sec::cryogenics}

The cryogenic system will consist of several sub-systems, which are required to initially liquefy the xenon target, maintain its constant low temperature during operation ($\Delta T/T < 0.05$\%), store it during down-time, and cleanse it of electronegative impurities which affect the light and charge yields, as well as from radioactive backgrounds. On-line measurements of the xenon purity will be performed by various diagnostic systems.

\paragraph{Cooling system} The cooling system will have to maintain a stable cryogenic environment for many years. It can be accomplished by means of cryocoolers, such as pulse tube refrigerators (PTRs)~\cite{ref::ptr} used in XENON, XMASS and PandaX, or by liquid nitrogen cooling~\cite{Bradley:2012fsa}, as chosen for LUX/LZ.
The design of the vacuum-insulated cryostat will be optimised in order to avoid heat leaks, and it will be equipped with super-insulation radiation shields. Given that the surface-to-volume ratio improves with the detector size, the cooling power required to keep the target cold will only be a few hundred Watts. Assuming similar cryogenics as in XENON1T, several redundant PTR cold-heads will be installed far away from the target volume, outside of the water shield. Xenon gas evaporated from the liquid target will be liquefied there, collected and returned into the main cryostat.

\paragraph{Storage} 
The storage of the noble liquid inventory of a multi-ton liquid xenon detector requires a dedicated solution. In addition, the purity must be maintained, thus continuous storage in a closed system is essential. The system must also be able to store and purify the noble gas before the detector is available, in order to considerably shorten the time required for detector filling as well as for initial gas purification. It should allow for quick recuperation of the noble gas in liquid form in case of emergency or in case of maintenance operations, as transfer to the gas phase and storage in bottles would take several weeks. The solution developed for XENON1T is a new storage system, {\it ReStoX}, which satisfies all these requirements~\cite{ref::restox_patent}.  {\it ReStoX} consists of a vacuum-insulated stainless steel sphere of 2.2\,m diameter, capable of holding 7.6\,tons of xenon in liquid phase. Xenon is kept in liquid form by means of a liquid nitrogen-based cooling system and can be constantly purified during storage. As the system is designed to withstand pressures of up to 72\,bar, the xenon can also be stored in gaseous phase at room temperature in case of longer shut-downs or emergency situations (longer power loss, etc.). Thus, a series of 7~interconnected {\it ReStoX} units is sufficient to store the full xenon inventory. 

\paragraph{Purification from electronegative impurities} 
To achieve a low energy threshold, good signal-background discrimination, and small signal corrections, the light and charge yields ought to be maximised. This requires a leak-tight detector and gas systems (metal seals) and constant purification of the noble-gas target from electronegative impurities such as N$_2$, O$_2$, H$_2$O, etc., which mainly stem from outgassing of material surfaces. During operation, the purification system constantly extracts gas from the detector and returns it after purification. While the MEG experiment~\cite{Adam:2013vqa} purifies its LXe target very efficiently in liquid form, this approach is not feasible for dark matter searches, due to the radioactive background induced by the recirculation pump for the liquid and due to radioactive contamination of the molecular sieves used for the cleaning. However, purification in the gas phase by means of hot zirconium getters at flow rates reaching up to $\sim$100\,slpm (corresponding to 880\,kg/day) has been achieved in the XENON1T R\&D programme, using effective heat exchangers~\cite{Giboni:2011wx}. Different recirculation pumps suitable for noble gases have been investigated, specifically addressing their performance in terms of gas purity. Since the entire xenon inventory will be in contact with the pump during the purification process, it must be leak tight and not emanate $^{222}$Rn. Standard diaphragm pumps, which were used for gas transport in previous experiments, do not fulfil the leak-tightness requirement. Thus, novel ultra-clean  magnetically driven piston pumps with hermetically sealed pumping volumes based on the design of~\cite{LePort:2011hy} are being developed and optimised. The performance of the purification system will be monitored online with dedicated purity monitors~\cite{Ferella:2007zza} and commercial systems, based on light absorption.

\paragraph{Purification from radioactive contaminants} 
Another aspect of purity is the cleanliness of the target with respect to radioactive noble gas isotopes, see also Section~\ref{sec::intrinsic}, which can easily mix with LXe.  Sub-ppt purification of xenon from krypton (necessary due to the radioactive isotope $^{85}$Kr) was already achieved \cite{Rosendahl:2015eor,Lindemann:2013kna} by an ultra-clean cryogenic distillation column developed within the XENON project~\cite{Rosendahl:2015eor,Rosendahl:2014wza,master_stolzenburg} together with a method to use the short-lived metastable $^\mathrm{83m}$Kr as a tracer to optimise the process \cite{Rosendahl:2015eor,Rosendahl:2014qqa,phd_rosendahl,master_stolzenburg}. Another problematic isotope is $^{222}$Rn, a daughter of the $^{238}$U chain which is present in the underground environment and also continuously emanated from the surfaces of the inner detector and cryostat itself. While metal seals mitigate the external $^{222}$Rn completely, all materials which are in contact with the target (liquid or gas) have to be specially selected for low Rn-emanation~\cite{Rau:2000,Kiko:2001}. Removing their topmost surface layer by electropolishing or etching can help to reduce this contribution. $^{220}$Rn, a daughter in the $^{232}$Th chain, could in principle also lead to backgrounds. However, due to its shorter half-life and lower abundance, it is less problematic. In addition, all efforts to reduce $^{222}$Rn are expected to also work for $^{220}$Rn.

Due to the minute amounts of radioactive impurities, separate online monitoring of the contamination level is not possible. Instead, the science data must constantly be checked for characteristic signatures of the contaminants. This is possible while the search regions for new physics remain blind. $^{222}$Rn ($^{220}$Rn) can be tagged efficiently using high-energy $\alpha$-events or the delayed $\beta-\alpha$ coincidence from the $^{214}$Bi-$^{214}$Po ($^{212}$Bi-$^{212}$Po) decay, and $^{85}$Kr can be identified using delayed $\beta-\gamma$ coincidence in $^{85}$Kr~$\to$~$^{85\rm{m}}$Rb~$\to$~$^{85}$Rb. Due to the low branching ratio of 0.454\%, the latter signature has a low efficiency and is thus not practical for low concentrations. For this reason, several off-line diagnostic methods were developed such as rare gas mass spectrometry (RGMS)~\cite{Lindemann:2013kna,master_stolzenburg} and atom trap trace analysis (ATTA)~\cite{Aprile:2013aza}. These have sensitivities below the 1\,ppt level and detect $^{\rm{nat}}$Kr rather than the radioactive isotope $^{85}$Kr itself, which is present at the $2 \times 10^{-11}$ level in $^{\rm{nat}}$Kr. A less precise (40\,ppt) but on-site and approximately on-line method using a quadrupole mass spectrometer following a cold-trap has been established as well~\cite{Dobi:2011vc,Brown:2012eb}.

\subsection{Signal readout}\label{sec:light}  

The sensitivity of liquid xenon detectors to dark matter is closely related to the detector's light collection and detection efficiency, as the expected WIMP scattering spectra show an exponential rise towards low nuclear recoil energies. As also explained in Section~\ref{sec::resolution},  the relevant quantity to be optimised is the light yield~(LY), which depends on the photon detection efficiency~(PDE) of the photosensors and on the detector's light collection efficiency; the latter is determined by the photocathode and light-reflector coverage, the transparency of the TPC electrodes, the reflectivity of the TPC walls, and the VUV photon absorption length. The detector configurations (single- or dual-phase), with their different geometrical photon detector coverage (4$\pi$ for single-phase detectors, top and bottom in present dual-phase detectors),  have a large impact on the light yield as well. This is because a large fraction of the emitted light can be absorbed if multiple reflections of VUV photons are required to reach the sensors. 

A common reference for the light yield is the detector response to the full absorption of 122\,keV gamma rays at zero drift field: XENON100 has a light yield of 4.3\,PE/keV~\cite{Aprile:2011dd}, LUX, which has superior PMTs and a higher PTFE reflectivity, reaches 8.8\,PE/keV~\cite{Akerib:2013tjd}, and the single-phase XMASS detector, where 62\% of the spherical surface is covered by PMTs with a QE of~28\%, has reached 14.7\,PE/keV~\cite{Abe:2013tc}. In the following, we review the relevant aspects for the light readout as studied within the DARWIN project, in order to achieve a design light yield around 8\,PE/keV.

\subsubsection{Photomultipliers}

At the operating cryogenic temperatures, standard bialkali photocathodes used in PMTs can have extremely low saturation currents due to the increase in resistivity of the photosensitive material. This problem was solved with Hamamatsu's Bialkali LT (Low Temperature) photocathode, which operates down to liquid nitrogen temperatures with high saturation current and high QE. This photocathode is available with 3''~diameter metal bulb photomultipliers. The Hamamatsu R11410 PMT with 12\,dynode stages, optimised for use in liquid xenon with a mean QE of 35\% at 178\,nm, reaching up to $\sim$40\% in some cases, and a collection efficiency (CE) of $>$90\%, has been investigated in setups relevant for the next-generation dark-matter search experiments using LXe~\cite{Baudis:2013xva,Lung:2012pi}. The results show a stable gain of $\sim$$5\times10^6$ at $\sim$1500\,V and an excellent peak-to-valley ratio around~3 or higher. A subset has been operated continuously in LXe for several months and repeated cooling cycles from room temperature to LXe temperatures were performed successfully, without damage to the PMTs and without changes in their response. The PMTs underwent several successful high-voltage tests in xenon gas and in strong electric fields~\cite{Baudis:2013xva}. After several iterations, and  collaborative efforts of Hamamatsu with  XENON, the radioactivity of these 3''~tubes was reduced to levels of $<$0.4\,mBq/unit from $^{238}$U,  0.02\,mBq/unit from $^{226}$Ra, and 0.01\,mBq/unit from $^{228}$Th~\cite{Aprile:2015lha}.

We note that the high QE of bialkali PMTs comprises a significant contribution ($\sim$20\%) of events where a single VUV photon releases two photoelectrons from the photocathode~\cite{Faham:2015kqa}. Thus, for a PMT with QE=35\% and CE=90\%, the PDE for photons impinging on the photocathode is $\sim$25\% (i.e., $\sim$80\% of the product QE$\times$CE). With $\sim$55\% of the instrumented area covered by a photocathode (fill factor), the overall PDE for a densely-packed PMT array is thus $\sim$14\% for state-of-the-art PMTs. However, this number is increased if the area between the PMTs is covered by efficient VUV-light reflectors.

\subsubsection{Novel photosensors}

PMTs, even if they are still the only photosensors used in current noble liquid dark matter detectors, have several important shortcomings: the residual radioactivity levels (although less relevant for very large detectors), cost, bulkiness, and stability at cryogenic conditions. Thus several alternative technologies are under consideration for DARWIN. The final choice of photosensors will be made during the design phase of the project, based on the technological maturity, performance characteristics (PDE, dark count rate, stability), radiopurity and cost of the different alternatives.

\paragraph{SiPM}
The silicon photomultiplier (SiPM) technology is rapidly developing and may become viable for readout of large detectors, offering very low radioactivity levels, compact geometry and low operation voltages. SiPMs may allow us to increase the photosensitive area coverage of the TPC and could in principle be suitable for 4$\pi$ coverage. Arrays of SiPMs sensitive to visible light and suitable for operation in LAr are becoming commercially available on surface-mount boards with >75\% fill factors over a 50$\times$50\,mm$^2$ area and are considered for the DarkSide-20k experiment~\cite{ref::ds_20k}. Several vendors are also developing VUV-sensitive SiPMs suitable for operation in LXe, and PDE values >10\% at 175-178\,nm have been reported~\cite{Ostrovskiy:2015oja}. However, photon detection in large-volume dark matter LXe detectors by means of SiPM arrays still requires significant optimisation. In particular, their present best dark count rate ($\gtrsim$1\,Hz/mm$^2$ at LXe temperatures~\cite{Ostrovskiy:2015oja}), should be reduced by $\sim$2 orders of magnitude to keep accidental coincidence low enough for the desired detection thresholds. Improvements are further required in the array PDE and correlated noise. The total correlated noise probability, namely the after-pulse and cross-talk probability, requires further studies as well.

\paragraph{SiGHT} 
Another solution could come from the development of novel vacuum photosensors with reduced radioactivity and a simpler internal structure than a multi-stage PMT. A new photosensor concept, the Silicon Geiger Hybrid Tube (SiGHT), is under development~\cite{ref::rossi2014}.  SiGHT, a descendant of the QUPID detector~\cite{Teymourian:2011rs}, consists of a 3''\,diameter cylindrical tube with a hemispherical photocathode biased at $-3$\,kV.  A SiPM is placed on a pillar within the cylinder. The entire structure is made from ultra-clean synthetic fused silica to ensure very low levels of radioactivity.  Electrons released from the photocathode are focused and accelerated onto the SiPM. A photoelectron hits a single pixel of the SiPM, yielding an output signal analogous to a single-photon hit on the SiPM.  The SiPM pixels allow for excellent charge resolution, which translates to the device's ability to count integer numbers of photoelectrons.   Good linearity can be achieved by using SiPMs with a high number of pixels.  The SiGHT photosensor was designed to operate at low temperatures, expecting dark count rates of the same order as PMTs at LXe temperatures (tens of Hz per tube).  Although the diameter of the photosensor currently being developed is 3'', future developments may include larger 4'' or 5''~versions.  The SiGHT photosensor could thus be well suited for ton- and multi-ton scale direct dark matter detection experiments.

\paragraph{Gaseous Photomultipliers}
Cryogenic Gaseous Photomultipliers (GPMs)~\cite{ref::chechik2008} could  become an economic alternative to PMTs for DARWIN, offering superior spatial resolution, compact geometry and similar overall PDE. GPMs combine a high-QE CsI-photocathode and cascaded gas-avalanche multipliers, e.g, Gas Electron Multipliers (GEMs) ~\cite{ref::gem} or Thick Gas Electron Multipliers (THGEMs) ~\cite{Breskin:2008cb})  coupled to an anode segmented into small pixels. Modular units, with a typical size of $\gtrsim$$20\times20$\,cm$^2$, equipped with UV-windows and embedded readout electronics, can be shaped to provide filling factors of $\sim$90\% (compared to present $\sim$55\% with circular PMTs), with comparable low radioactivity. For a nominal QE of $\sim$25\% at 175\,nm~\cite{ref::Qe_CsI} and optimised choice of hole geometry and counting gas (Ne/CH$_4$ or Ar/CH$_4$ mixtures), an overall PDE of $\sim$15\% over the entire instrumented area can be expected, similar to PMT arrays. While the superior granularity is not an {\sl a priori} critical requirement, it may prove to be useful for precise event topology reconstruction.

The electron multipliers would be either cascaded THGEMs or hybrid structures, e.g., CsI-coated THGEM followed by thin-mesh multipliers~\cite{ref::duval1,Duval:2011ky} with high gains ($>$10$^5$), allowing for high single-photon detection efficiency at LXe temperatures. Such GPMs, with reflective CsI photocathode, are suitable for the top photosensor array. GPMs could be also deployed at the TPC walls to provide $\sim$4$\pi$ coverage with a considerable improvement in the S1 photon detection due to the reduction in multiple reflections (the degree of improvement will depend on the ratio between the absorption length and TPC diameter). The wall GPM should include an additional semi-transparent photocathode, on the inner window surface, to prevent loss of photons by total internal reflection. Stable operation of a 4''~triple-THGEM GPM with a reflective CsI photocathode, coupled to a dual-phase LXe TPC, was recently demonstrated. It displayed a broad dynamic range, namely the capability to detect both single photons and massive $\alpha$-induced S2~signals under the same conditions~\cite{Arazi:2015pta}. The feasibility of the  $4\pi$ concept and the GPM PDE optimisation are subject to ongoing studies.

\subsubsection{Liquid Hole-Multipliers: charge and light readout in a single-phase TPC}

\begin{figure}[b!]
\begin{center}
\includegraphics*[width=0.8\textwidth]{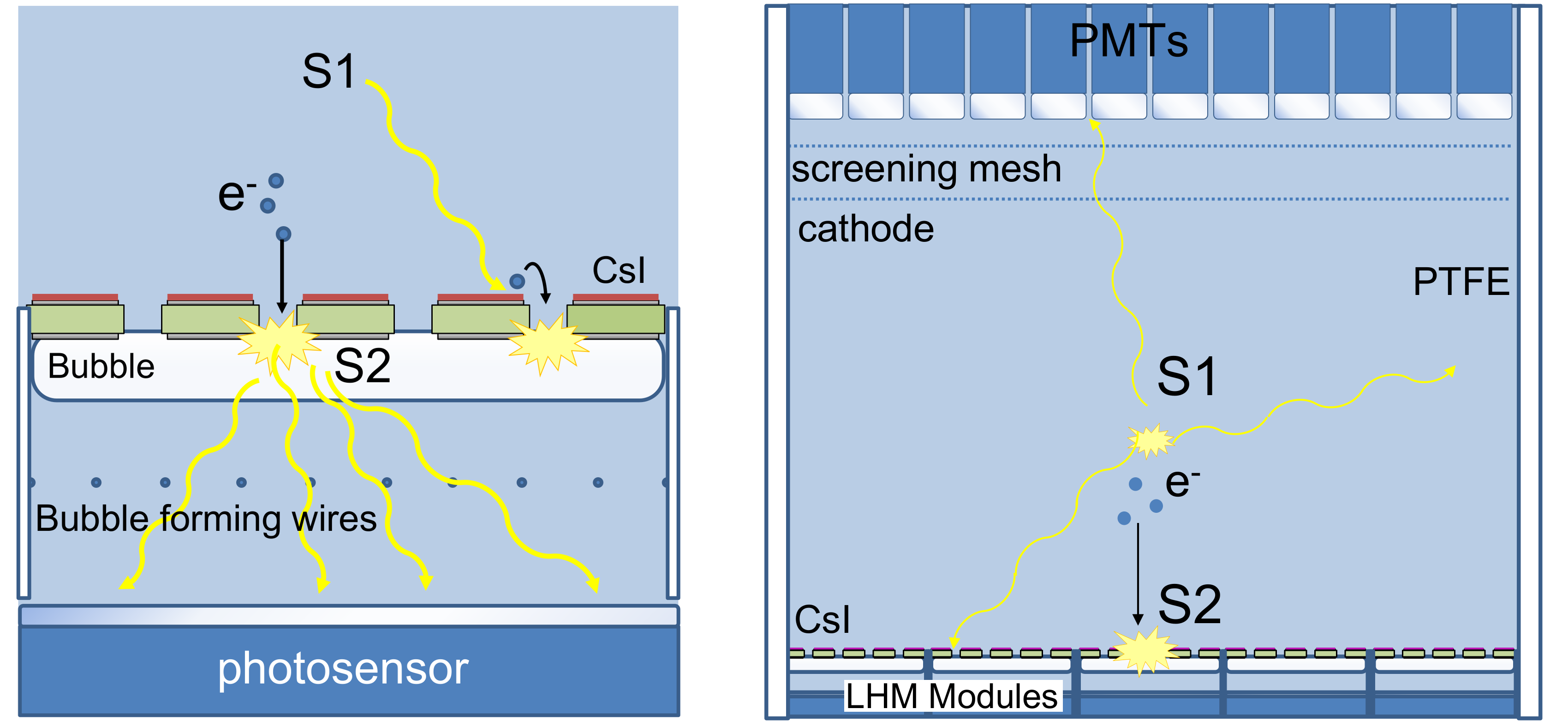} 
\end{center}
\caption{The liquid hole multiplier concept: {\bf (left)} A single LHM element, coated with CsI to allow for the detection of S1 photons in addition to ionisation electrons. Controlled bubble	formation can be achieved by ohmic heating of resistive wires below the electrode. {\bf (right)} Schematic design of a liquid-only single-phase TPC with bubble-assisted LHMs at the bottom. S2 signals are created by electrons drifting down towards the LHM. Accurate S2-based position reconstruction is permitted by an array of photosensors below the LHMs. Figure adapted from~\cite{Arazi:2015uja}. \label{fig::lhm}}
\end{figure}

The Liquid Hole-Multiplier (LHM) is a new, ``non-traditional'', detection concept~\cite{Breskin:2013ifa}. It originated from the concern that it may be quite difficult to maintain the liquid-gas interface, anode and gate completely parallel to each other across the diameter of a large dual-phase TPC; this may result in degraded S2~resolution and, as a possible consequence, reduced S2/S1-based background discrimination capability. This problem could be solved, in principle, if S2~light were to be generated in a high-field region within the liquid itself rather than in the vapour phase~\cite{Giboni:2014tqa}. Some groups studied the possibility of generating S2 light around thin wires immersed in LXe~\cite{Aprile:2014ila,Giboni:2014tqa,Ye:2014gga}, following works from the 1970s~\cite{Doke:1982tb}. Parallel efforts focus on immersed THGEM and GEM electrodes. First experiments with a THGEM immersed in LXe demonstrated large S2 signals for alpha-particle induced ionisation electrons~\cite{Arazi:2013fxa}. Subsequent studies~\cite{Arazi:2015uja,Erdal:2015kxa} proved that the light was in fact generated at the bottom part of the THGEM hole, in a xenon gas bubble trapped below the electrode, see Figure~\ref{fig::lhm}~(left). It was shown that the process can be controlled and maintained stable over many days. An S2~resolution ($\sigma/E$) of $\sim$7.5\% was demonstrated for $\sim$6000~ionisation electrons, significantly better than in XENON100 ($\sim$10-12\%)~\cite{Aprile:2011dd}. Similar results were reached with immersed GEM electrodes~\cite{Araz:2016kxa}. With an appropriate field configuration the process can yield up to a few hundred photons per electron. Coating the electrode with CsI permits the detection of S1 photons in addition to ionisation electrons for the same field configuration, with a drift field of $\sim$0.5--1 kV/cm in the liquid~\cite{Erdal2016}. Immersed GEMs are preferable over THGEMs for photon detection, as they allow for higher electric fields at the CsI~photocathode surface and thus better photoelectron extraction. Based on a previous study, which indicated that the QE of CsI immersed in LXe is $\sim$30\% at 175\,nm for a sufficiently large field on the photocathode surface~\cite{Aprile:1994bw}, the PDE across the CsI-coated GEM electrode is expected to be $>$15\%.

Looking forward, GEM-based LHM modules with individual heating elements to generate the trapped bubbles, may be tiled to form a large instrumented surface sensitive to both ionisation electrons and S1 photons. This configuration, with an inverted drift field, is shown schematically in Figure~\ref{fig::lhm}~(right). It has two potential merits: first, it allows for a high S2 resolution over the entire area, by local control of the liquid-gas interface inside the module; second, it can considerably boost the TPC's light yield by the removal of three grids (gate, anode and screening electrode) as well as the liquid-gas interface itself. Even if the absorption length is not much larger than the TPC diameter, the loss of S1 photons is prevented by the effective elimination of multiple reflections: photons will typically reflect only once off the TPC PTFE wall before being detected by either the PMT array or the LHM array. Light readout of such LHM modules can be performed by PMTs, SiPMs or GPMs (the latter with reflective CsI photocathodes); the high dark count rate of SiPMs will be of no concern in this case as each primary S1 photon will produce a `flash' of light comprising dozens of secondary photons inside the LHM module. While the basic LHM configuration comprises a bubble trapped below the electrode (and thus requires using LHMs horizontally at the bottom of the TPC), ongoing studies look into the possibility of trapping a gas layer above the electrode -- opening the possibility of using LHMs at the top of the TPC.

\subsection{Calibration} 
\label{sec::calibration}

The signal response of liquid xenon detectors to particle energy deposits in the active volume are determined through regular calibration campaigns. The first generation of experiments has mainly relied on the use of external $\gamma$-sources (such as $^{137}$Cs, $^{60}$Co, $^{228}$Th) and broad-band ($^{241}$AmBe, $^{252}$Cf) or mono-energetic (generators) neutron sources for detector calibration. The considerably larger ton-scale detectors and  in particular the multi-ton instrument DARWIN render calibration more challenging, as the effective background reduction by self-shielding also suppresses the detection of the majority of particles from external sources. It is however the innermost, central detector region which is used for new physics searches and thus must be calibrated precisely. 

Several groups have started to investigate the use of sources dissolved in the liquid target, so-called internal sources. Only isotopes with a very short half-life can be employed for regular calibrations, followed by science runs.  Some of these sources are neutron-activated xenon, providing rather high-energy $\gamma$-lines at 164\,keV ($^{131\rm{m}}$Xe, $T_{1/2}=11.8$\,d) and 236\,keV ($^{129\rm{m}}$Xe, $T_{1/2}=8.9$\,d)~\cite{Ni:2007ih}.  However, the half-lives of the activated Xe isotopes are too long, and their energies too high, to be useful for ton-scale detectors.

\begin{figure}[b!]
\begin{minipage}[]{0.48\textwidth}
\begin{center}
\includegraphics*[width=1.\textwidth]{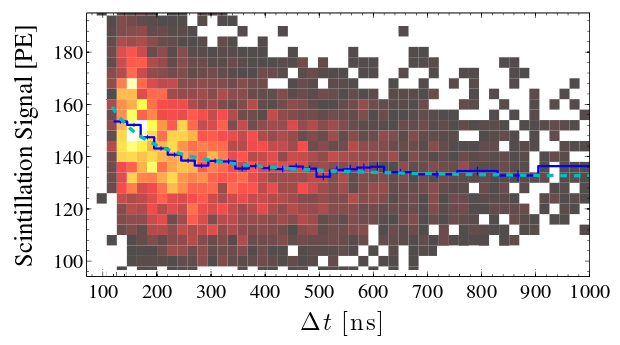}  
\end{center}
\end{minipage}
\hfill
\begin{minipage}[]{0.52\textwidth}
\vspace{-0.9cm}
\caption{Scintillation light from the 9.4\,keV line of $^{83\rm{m}}$Kr, following the initial 32.1\,keV $\gamma$-ray. The half-life of the intermediate state is 154\,ns. The light signal depends on the time difference $\Delta t$ between the two events, due to the presence of ionisation from the first interaction. Therefore only 9.4\,keV events with large $\Delta t$ can be used for calibrations. Figure adapted from~\cite{Baudis:2013cca}. \label{fig::krline} }
\end{minipage}
\end{figure}

The best-studied example of a  short-lived internal source is $^{83\rm{m}}$Kr~\cite{Kastens:2009pa,Manalaysay:2009yq}, a daughter of $^{83}$Rb. It has been demonstrated that no long-lived Rb-isotopes were emitted into a $^{83\rm{m}}$Kr-sample for calibration~\cite{Hannen:2011mr}. $^{83\rm{m}}$Kr has a half-life of 1.83\,h, and decays via 32.1\,keV and 9.4\,keV conversion electrons, where the intermediate state has a lifetime of 154\,ns. The second low-energy process takes place close to the first one and has an increased light signal due to an additional supply of ions and electrons that failed to recombine from the preceding transition. This limits the use of the 9.4\,keV line to the tail of the time distribution where the effect becomes irrelevant~\cite{Baudis:2013cca}, see also Figure~\ref{fig::krline}. $^{83\rm{m}}$Kr has been successfully used in DarkSide-50~\cite{Agnes:2014bvk}, XENON100 and LUX~\cite{Akerib:2013tjd}. 

The LUX collaboration has also successfully employed tritiated methane (CH$_3$T) as an internal calibration source~\cite{Akerib:2015wdi}. With $T_{1/2}=12.3$\,y, the half-life of tritium is too long to leave it to decay in the detector, however, the collaboration has demonstrated that it can be efficiently removed by a hot zirconium getter~\cite{Dobi:2010ai}, designed to remove gases (N$_2$, O$_2$, H$_2$O, etc.) and other electronegative impurities from the liquid noble gas target (see Section~\ref{sec::cryogenics}).
The XENON collaboration has used tritiated methane in XENON100, and is currently performing R\&D towards the use of $^{220}$Rn, emanated from a $^{228}$Th source~\cite{Lang:2016zde}. The use of the same source was also studied by XMASS~\cite{Kobayashi:2016vwj}. It exploits the short half-life of $^{220}$Rn and the subsequent $\beta$-decays of its daughters $^{212}$Pb and $^{208}$Tl, which are in radioactive equilibrium with the rest of the chain.

The light and charge response of dual-phase TPCs for dark matter searches must be calibrated for various reasons:

\paragraph{Energy calibration} The use of the relative scintillation efficiency $\leff(E_{\rm{nr}})$ and the charge yield $Q_y(E_{\rm{nr}})$, which describe the size of the signal from a nuclear recoil of energy $E_{\rm{nr}}$ with respect to a fixed-energy $\gamma$-source, requires a calibration of the detector with mono-energetic $\gamma$-lines. The traditionally used response to 122\,keV $\gamma$'s from a $^{57}$Co source was recently replaced by the intrinsic standard 32.1\,keV line from $^{83\rm{m}}$Kr. This single line allows for the light/charge calibration of a DARWIN detector using the relative efficiencies measured in dedicated experimental setups, under tightly controlled conditions, as elaborated in Section~\ref{sec::ly}. 

For the characterisation of the electronic recoil background up to several MeV, an energy scale set by the full absorption peaks of various $\gamma$-sources is desirable. Intrinsic sources are mandatory to populate the inner detector regions with a sufficient number of events. One possibility is using the various decay lines seen in $^{220}$Rn and its daughters~\cite{Lang:2016zde}. Another possibility is the known shape of $\beta$-spectra from intrinsic contaminations and sources, e.g., the $2\nu\beta\beta$-spectrum from $^{136}$Xe. 
 
\paragraph{Nuclear recoil band (signal-like events)} The detector response to single-scatter nuclear recoils down to lowest energies ($\sim$1\,keV$_{\rm{nr}}$) must be known precisely. Neutron-induced recoil spectra generated by broad-spectrum sources such as $^{241}$AmBe and $^{252}$Cf, or by mono-energetic MeV-neutrons from generators, are representative for  WIMP-generated spectra of $m_\chi\gtrsim50$\,GeV/$c^2$. The correct light and charge distribution for lower-mass WIMPs is to be determined by feeding the results from these calibrations into Monte Carlo codes, which take into account the signal resolutions~\cite{Aprile:2013teh,Sorensen:2012ts}. 

A direct calibration of the signal distribution with external neutron sources, preferentially mono-energetic neutrons from a generator in order to exploit the kinematics of multiple-scatter processes, is still feasible for DARWIN, thanks to the rather long mean-free path of neutrons in LXe. A direct calibration of the low-energy nuclear recoil response with an $^{88}$YBe-source, as recently proposed by~\cite{Collar:2013xva} and currently studied in XENON100, will be only possible in much smaller experimental setups.
 
\paragraph{Electronic recoil band (background-like events)} To establish a background model for the WIMP dark matter search,  the distribution of the ER signal in the TPC must be well understood. The same holds for all searches using the ER signal itself, such as solar neutrinos and axions, see Section~\ref{sec::other}. This background is mostly due to target-intrinsic $\beta$-contaminations.  Line sources, such as $^{83\rm{m}}$Kr, cannot be used for calibration as a continuous spectrum is required. We thus expect to use intrinsic sources such as tritiated-methane and $^{220}$Rn. These provide a continuous spectrum of low energy recoils through ground-state to ground-state beta decays either directly (TCH$_{3}$) or through their daughters ($^{220}$Rn). The comparison of signal and background calibration samples allows for the determination of the background discrimination level. The charge-to-light ratio exploits the different (position-corrected and energy-dependent) mean values of the S2/S1-ratio for signal and background, thanks to the different energy loss mechanisms of the recoils~\cite{ref::phd_dahl}.

\subsection{Light and charge yield of electronic and nuclear recoils}
\label{sec::ly}

The response of  LXe to particle interactions with energy depositions around the detection threshold is of high relevance, as the differential nuclear recoil spectrum induced by WIMP-nucleus elastic scattering is exponentially decreasing. In particular, WIMPs with masses below 10\,GeV/$c^2$ could potentially leave signatures only in the lowest energy bins of a liquid xenon detector. Such measurements have been performed within the DARWIN consortium in the last years~\cite{Aprile:2008rc,Plante:2011hw,Aprile:2013teh} and new measurements are currently ongoing.

The response of large LXe detectors to low-energy nuclear recoils can be measured {\it {in situ}} using monoenergetic neutrons from deuterium-deuterium (D-D) fusion generators, as shown by LUX~\cite{Akerib:2015rjg} and planned for the XENON1T experiment. Such a measurement, in combination with an empirical response model that is fitted to the data, allows for the determination of the charge and light yields down to 1\,keV$_\textnormal{nr}$ energy, or even below~\cite{Akerib:2015rjg}.

To date, estimates of the dark matter sensitivity of LXe TPCs assume that electric fields have a small effect on the light yield from nuclear recoils. This assumption is supported by initial measurements~\cite{Manzur:2009hp}, indirect analyses~\cite{Aprile:2013teh} as well as semi-empirical models (NEST)~\cite{Szydagis:2011tk,Szydagis:2013sih}. Direct measurements in small LXe detectors are nonetheless ongoing.

For standard WIMP searches, electronic recoils are expected to stem from background events, however, for axion searches, leptophilic dark matter models or low-energy neutrino measurements, ERs turn into the expected signal (see also Sections~\ref{sec:axions} and~\ref{sec:solar}). The scintillation yields of ERs in liquid xenon were measured with small cryogenic cells using Compton-scattered photons from collimated, high-activity $^{137}$Cs sources in coincidence with HPGe and NaI detectors placed under various scattering angles, corresponding to energies down to 1.5\,keV~\cite{Aprile:2012an,Baudis:2013cca}.  Results from data at zero electric field show a decrease of the yield of recoiling electrons below 20\,keV, to a level of $\sim$~40\% of its value at higher energies at around 1.5\,keV. Measurements of the light quenching in an electric field have also been performed~\cite{Baudis:2013cca}. The results have been used to set an energy scale for axion searches with XENON100~\cite{Aprile:2014eoa}, to test leptophilic dark matter models~\cite{Aprile:2015ade}, and for the first search for an annual modulation signal with a LXe TPC~\cite{Aprile:2015ibr}.

\subsection{Detector resolution} 
\label{sec::resolution}

The light and charge signals are both employed in the data analysis process. While the energy scale can be derived either using one of these~\cite{Aprile:2012nq} or from their linear combination~\cite{Akerib:2013tjd}, background discrimination via the S2/S1~ratio always requires the precise knowledge of both quantities. The smaller and hence more-difficult-to-detect~S1 signal sets the energy threshold of the detector. The energy resolution is derived from the individual signals or their combination, and is related to the number of detected physical quanta. For the S1~channel, this is the number of photons recorded by the photosensors, giving rise to signals measured in PE. Because of the finite quantum efficiency ($\sim$35\% for state-of-the-art PMTs), photoelectron collection efficiency ($\sim$90\%) and light collection efficiency (LCE), which describes the fraction of primary photons reaching a photosensitive area, the detected number of PE is considerably smaller than the number of initially generated photons. The LCE depends on the target purity, the reflectivity of the inner TPC surfaces,  the transparency of the TPC electrodes, the photocathode coverage of the detector, as well as on the TPC height-to-diameter ratio. All these parameters can be optimised for a given detector. The light yield of the LUX detector, 8.8\,PE/keV$_\textnormal{ee}$ for a 122\,keV line (electron recoil equivalent) at zero-field~\cite{Akerib:2013tjd}, is about 2$\times$ higher than the one of XENON100 (4.3\,PE/keV$_\textnormal{ee}$~\cite{Aprile:2011dd}), mainly due to an improved PTFE reflectivity, optimised TPC electrodes transparency and a higher collection efficiency of the PMTs. Typically, PTFE is used as an efficient reflector for the scintillation light of LXe at 178\,nm, with a reflectivity above 90\%~\cite{ref::ptfe_reflectivity}. 

\begin{figure}[b!]
\begin{minipage}[]{0.49\textwidth}
\begin{center}
\includegraphics*[width=1.\textwidth]{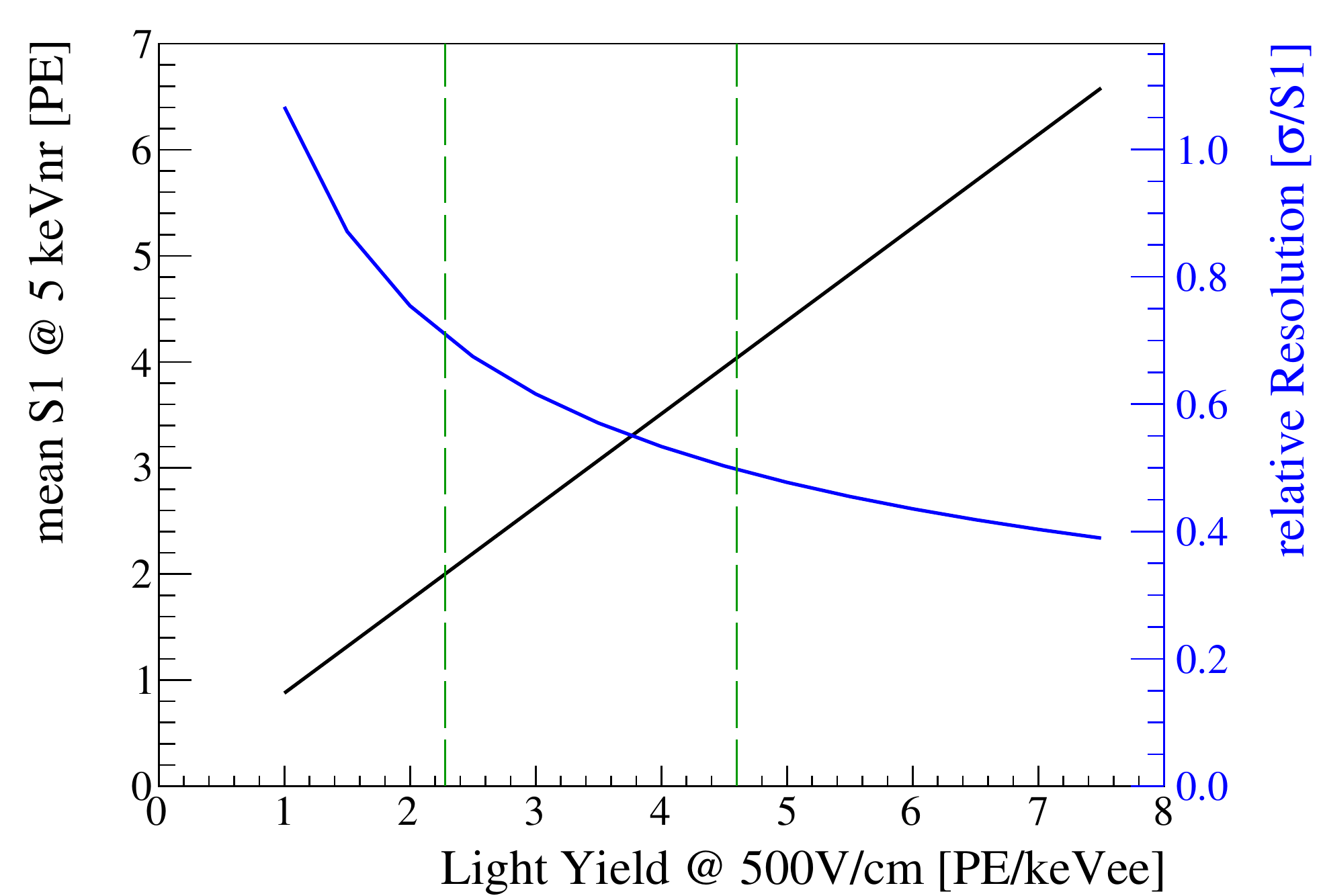}  
\end{center}
\end{minipage}
\hfill
\begin{minipage}[]{0.49\textwidth}
\begin{center}
\includegraphics*[width=1.\textwidth]{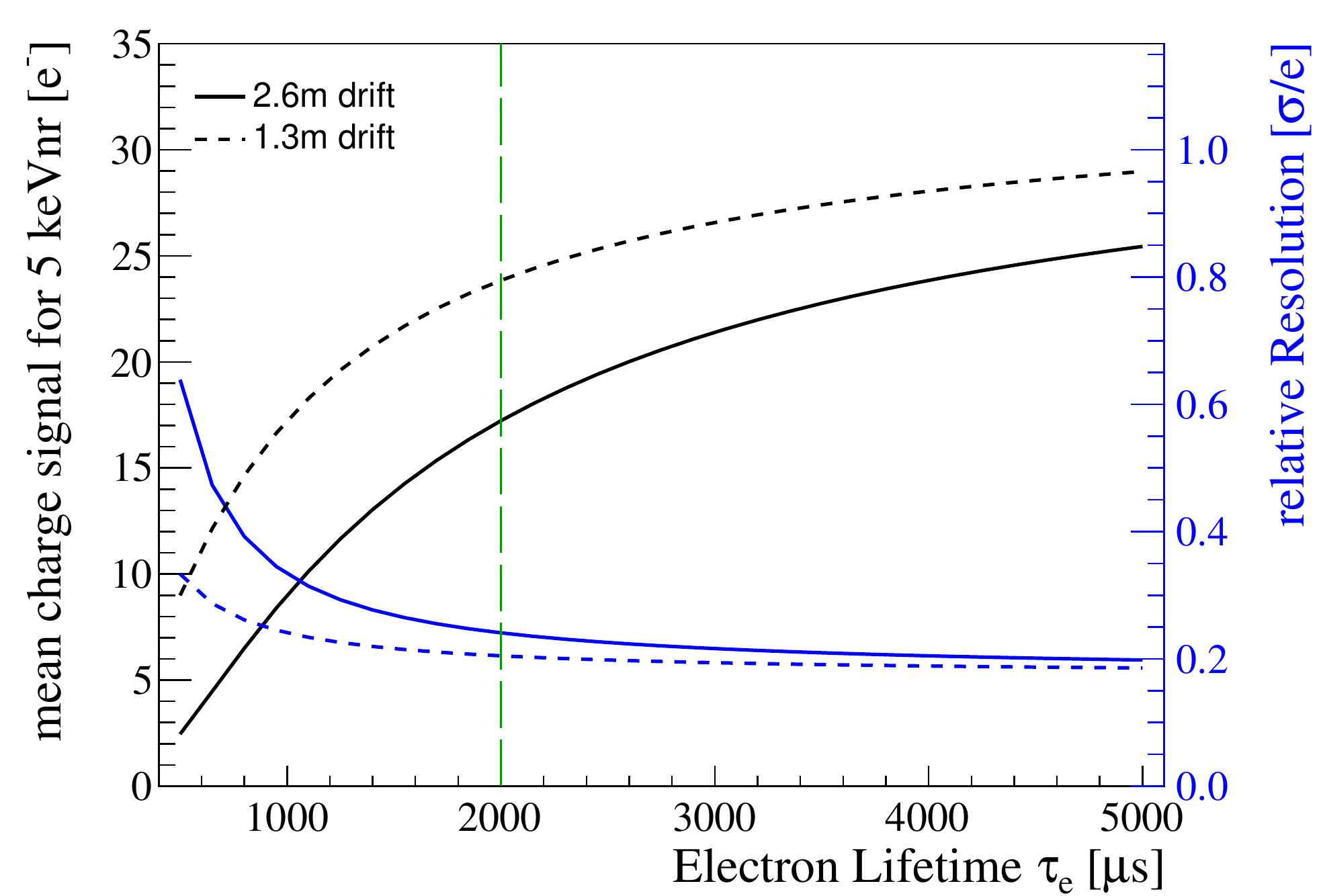}  
\end{center}
\end{minipage}
\caption{{\bf (left)} Mean size of the S1 light signal in a dual-phase LXe TPC, expected from a nuclear recoil of 5\,keV$_\textnormal{\tiny nr}$ (left axis), plotted against the detector's light yield for a 122\,keV $\gamma$-line at a drift field of $\sim$500\,V/cm. The relative scintillation efficiency \leff \ of LXe for this study is taken from~\cite{Aprile:2011hi}. Also shown is the relative resolution ($\sigma$/S1) at 5\,keV$_\textnormal{\tiny nr}$, assuming that it is dominated by a Poisson process (right axis). The LXe TPCs XENON100~\cite{Aprile:2011dd} and LUX~\cite{Akerib:2012ak} have achieved light yields of $\sim$2.3\,PE/keV$_\textnormal{\tiny ee}$ and $\sim$4.6\,PE/keV$_\textnormal{\tiny ee}$ at this drift field, respectively, as indicated by the green lines. {\bf (right)} Mean charge signal in electrons, before gas amplification, for a 5\,keV$_\textnormal{\tiny nr}$ recoil signal as a function of the electron lifetime $\tau_e$. A drift field of $\sim$500\,V/cm is assumed. The charge yield $Q_y$ for LXe is taken from~\cite{Aprile:2013teh}. The right axis shows the relative Gaussian resolution for 1.3\,m and 2.6\,m electron drift, corresponding to the central and the maximal value in a DARWIN detector. Lifetimes of 2\,ms have already been achieved in LXe detectors (green line). In both cases, it is unrealistic that DARWIN can improve significantly with respect to the numbers already achieved, leading to S1 resolutions of about~40\% for 5\,keV$_\textnormal{\tiny nr}$ recoil signals and to charge resolutions of about~20\%.  \label{fig::resolution} }
\end{figure}

Due to electron-ion recombination effects, the light yield (LY) is always highest for zero-field and decreases (in an energy-dependent way) with increasing drift field, see for example~\cite{Manalaysay:2009yq}. The numbers for XENON100 and LUX quoted above reduce to 2.3\,PE/keV$_\textnormal{ee}$ and $\sim$4.6\,PE/keV$_\textnormal{ee}$ at 122\,keV$_\textnormal{ee}$ and $|\vec{E}|\sim0.5$\,kV/cm, respectively. Figure~\ref{fig::resolution}~(left) illustrates this situation for a nuclear recoil signal of 5\,keV$_\textnormal{nr}$: a higher light yield will improve the S1-resolution at a given nuclear recoil energy $E_\textnormal{nr}$. However, current dark matter TPCs are already highly optimised, hence not much improvement beyond values of $4-5$\,PE/keV$_\textnormal{ee}$ at a 0.5\,kV/cm drift field (about $8-10$\,PE/keV$_\textnormal{ee}$ at zero field) can be realistically expected for DARWIN, leading to an anticipated S1-resolution of around~40\% at the detector threshold.

Due to the larger number of quanta involved, the resolution of the proportional S2 signal (from the ionisation electrons) is superior to the S1~resolution. The $W$-value, describing the energy required to create an electron-ion pair, is 15.6\,eV in xenon~\cite{Takahashi:1975}. The number has to be corrected for recombination effects in the non-zero electric field, leading to a somewhat larger effective $W$-value. However, the total number of electrons liberated by low-energy interactions is still rather high. As the typical electron extraction fields of $>$9\,kV/cm lead to  100\%~extraction efficiency~\cite{Aprile:2013blg}, the main loss mechanism is the effective electron lifetime, $\tau_e$, due to capture of electrons by electronegative impurities in the target. Electron lifetimes of 2\,ms  have been demonstrated in large LXe detectors~\cite{ref::xe1t_demonstrator} and its impact on the resolution is illustrated in Figure~\ref{fig::resolution}~(right): for $\tau_e\geq2$\,ms and 100\% extraction efficiency, the charge resolution is better than 24\% for 5\,keV$_\textnormal{nr}$ recoils throughout the TPC. For $\tau_e=5$\,ms, the resolution improves to 20\%. The number of photoelectrons detected by the PMT arrays is typically about~20\,PE/$e^-$~\cite{Aprile:2013blg}, and depends on the gas pressure, the distance between liquid-gas interface and anode as well as the extraction field. Due to the much larger number of quanta, the fluctuations in the number of PE are subdominant compared to the fluctuations in the number of electrons. 

The linear combination of light and charge signals, exploiting an anti-correlation between the two in ERs~\cite{Aprile:2006kx}, allows for energy resolutions similar to NaI~crystals  ($\sigma/E=1.53$\% has been reached at 2480\,keV~\cite{Albert:2014awa}). Such a resolution would be relevant for many of the non-WIMP searches which focus on  ERs, such as searches for neutrinoless double beta-decays or for axions and ALPs. For low-energy nuclear recoils, no strong anti-correlation is expected in LXe because the recombination-fluctuations are sub-dominant compared to the uncorrelated S1~and S2~fluctuations~\cite{ref::phd_dahl,Sorensen:2010hq}.
The resolution of a combined light and charge signal close to threshold is therefore simply given by the sum of the physical quanta. It is dominated by the charge signal and therefore only leads to a minor improvement of the resolution compared to an S2-only scale~\cite{Schumann:2015cpa}. As charge and light signals are both smaller for NRs compared to ERs, due to quenching effects, the energy resolution of an ER signal will always be superior to the one of a NR of the same energy, regardless of how the energy scale is reconstructed.

\subsection{Data acquisition and trigger schemes}\label{sec:daq}
 
This section addresses issues regarding the electronics and data acquisition (DAQ). While the total number of channels in DARWIN is still moderate compared to accelerator-based physics experiments, the small, keV-sized signal in the dark matter channel requires an extremely low energy threshold and excellent noise conditions. The digitized waveform of every channel is recorded in order to use the maximum amount of information for further data analysis. The DAQ system must also be able to handle the different data taking rates in science mode and during calibration runs.
 
The DAQ system of a large dual-phase TPC has to address several challenges. The number of channels, $\geq$1000 photosensors, depending on the sensor size, will be several times higher than in present (e.g., XENON100: 242~\cite{Aprile:2011dd}, LUX: 122~\cite{Akerib:2012ys}, PandaX-II: 110~\cite{Tan:2016diz}, XENON1T: 248~\cite{Aprile:2015uzo}) or upcoming (e.g., LZ: 488~\cite{Akerib:2015cja}) detectors. The maximum time difference between the S1 and S2 signals will be of the order of one millisecond.  The detailed waveforms, at $\cal O$($2-10$)\,ns resolution, must be digitised as they contain information for the S2/S1~background discrimination, noise rejection, etc. However, there is little  information stored in the comparatively long time between the S1 and S2 peaks and  already present-day detectors do not digitize this part of the waveforms to reduce the amount of data~\cite{Aprile:2011dd,Akerib:2012ys}.
 
The causal connection between S1 and S2~signals limits the maximum achievable detector rate due to the possibility of event pile-up. While conservatively high trigger rates of $\sim$10\,Hz in dark matter mode would only lead to $\sim$1.5\% of events suffering from pile-up in DARWIN, large LXe detectors with their excellent self-shielding capabilities will generally require very large amounts of calibration data to reach a sufficient number of calibration events within the fiducial volume, hence a high calibration rate, see Section~\ref{sec::calibration}. In a ``classical'' event-based readout using a common trigger, the events will overlap and a constant acquisition window will lead to information loss. Finally, the trigger threshold must be as low as possible to ensure a low energy threshold.
 
The increased number of channels in DARWIN can be handled through parallelisation,  the level of which can be increased according to the actual requirements of the experiment. All readout channels will operate independently from one another, they are not triggered globally but run in ``self-triggering''-mode. The data from all channels will be correlated and reconstructed in real-time on commodity computing hardware, which only keeps the information associated with an event for storage. Such a flexible software trigger can also be further parallelised in order to increase the computation speed. It will allow for detector calibration in pile-up mode, as overlapping events can be accepted while the correlation analysis of the S1 and S2~peaks is postponed. This  analysis could be based on the reconstructed $xy$-positions of the S1 and the S2 signals, as well as on the size of the S2/S1 ratio. 
 
Since the data is fully reconstructed before making a storage decision, more sophisticated filtering algorithms can be used to enhance DARWIN's physics capabilities. Examples are: (i) $xy$-position reconstruction to only store events reaching the central volume during calibration runs, (ii) a relatively small, random subset of events can be stored  to study high-energy and background events, and (iii) specialised triggers can be used to study specific  background topologies (e.g., delayed coincidence triggers to select $^{214}$Bi-Po events to study $^{222}$Rn backgrounds).

A DAQ system designed according to this concept was developed for the XENON1T detector. Because of its scalability by increasing the level of parallelisation, it can be regarded as the first step towards a DAQ system for DARWIN.

\section{Summary and Outlook}

DARWIN will be the ultimate liquid xenon dark matter detector with a sensitivity for spin-independent WIMP-nucleon cross sections down to $\sim$10$^{-49}$\,cm$^2$ capable to detect or exclude WIMPs with masses above $\sim$5\,GeV/$c^2$. With its large target mass, low energy threshold, and ultra-low background level, DARWIN  will also provide a unique opportunity for other rare event searches such as axions and other weakly interacting light particles. It will address open questions in neutrino physics, e.g., by measuring the low-energy solar neutrino spectrum with better than 1\% precision or by searching for the neutrinoless double beta decay of $^{136}$Xe. At its lowest energies, the DARWIN detector will provide the possibility to observe coherent neutrino-nucleus interactions from solar $^8$B neutrinos, to precisely test the standard solar model flux-prediction, and to detect neutrinos from galactic supernovae.

DARWIN will employ  a time projection chamber filled with 40\,t of liquid xenon;  the full instrument will require about 50\,t of liquid xenon.  A vigorous R\&D and design effort is ongoing within the international DARWIN consortium. It comprises technical aspects such as the design and prototyping of the time projection chamber, Monte Carlo studies of the expected radiogenic and cosmogenic backgrounds, investigation of new light and charge readout schemes and of novel sensors to operate in  liquid xenon, selection of low-background construction materials by means of high-purity germanium spectroscopy and other techniques, radon emanation measurement and removal, new data acquisition and trigger schemes, data analysis and also addresses the scientific reach of the facility. Charge and light yield measurements of nuclear and electronic recoils  at lowest energies,  necessary to define accurate energy scales in a LXe dark matter detector, are ongoing at several institutions. In parallel to the baseline dual-phase detector geometry, single-phase TPC concepts are being evaluated and prototyped. The R\&D and design phase will end by 2019, after which the construction of the various sub-systems will start. Following detector  installation and commissioning in the underground laboratory, a first science run could start by 2023. To fully exploit its WIMP sensitivity, the facility would be operated for at least 7\,years.

In summary, DARWIN has a  unique discovery potential in the areas of astroparticle and low-energy neutrino physics.

\appendix
 
\section*{Acknowledgments}
This work has been supported by the ASPERA first common call (EU), the University of Zurich (CH), the Albert Einstein Center at the University of Bern (CH), the Swiss National Foundation (SNF), the FP7 Marie Curie-ITN action {\it Invisibles} (EU), the Max-Planck Society (DE), the PRISMA Cluster of Excellence in Mainz (DE), the Helmholtz Alliance for Astroparticle Physics (DE), the Israel Science Foundation (IL), the MINERVA Foundation (DE), the Stichting Fundamenteel Onderzoek der Materie (NL), the Funda\c{c}\~{a}o para a Ci\^{e}ncia e Tecologia (PT), the Imperial College Trust and the Science and the Technology Facilities Council (UK),  the Istituto Nazionale di Fisica Nucleare (IT), and the National Science Foundation (USA).

\bibliographystyle{JHEP}
\bibliography{darwin_sensitivity_v6b}

\end{document}